\documentclass[12pt]{article}
\usepackage[margin=2cm]{geometry}
\usepackage{comment}
\usepackage{amsmath,amssymb,mathtools,
graphicx,subfigure,setspace,physics}
\usepackage{cite}
\usepackage{slashed} 		 
\usepackage{xcolor}
\usepackage{listings}
\usepackage{tcolorbox}
\usepackage{placeins}
\tcbuselibrary{breakable}

 \usepackage{url} 

\usepackage[utf8]{inputenc}

\newcommand{\be}{\begin{equation}}
\newcommand{\bea}{\begin{eqnarray}}
\newcommand{\eea}{\end{eqnarray}}
\newcommand{\ba}{\begin{array}}
\newcommand{\ea}{\end{array}}

\newcommand{\ee}{\end{equation}}
\newcommand{\bes}{\begin{equation*}}
\newcommand{\beas}{\begin{eqnarray*}}
\newcommand{\eeas}{\end{eqnarray*}}
\newcommand{\bas}{\begin{array*}}
\newcommand{\eas}{\end{array*}}
\newcommand{\ees}{\end{equation*}}

\numberwithin{equation}{section}
\begin{document}
\onehalfspacing
\noindent

\begin{titlepage}
\vspace{10mm}
\begin{flushright}
\end{flushright}

\vspace*{20mm}
\begin{center}

{\Large {\bf Eigenstate Thermalization Hypothesis: A Primer 
}\\
}

\vspace*{15mm}
\vspace*{1mm}
{Mohsen Alishahiha and  Mohammad Javad Vasli}

 \vspace*{1cm}

{\it { School of Quantum Physics and Matter\\ Institute for Research in Fundamental Sciences (IPM),\\
	P.O. Box 19395-5531, Tehran, Iran\\}  }

 \vspace*{0.5cm}
{E-mails: {\tt alishah@ipm.ir, vasli@ipm.ir}}%

\vspace*{1cm}
\end{center}

\begin{abstract}
How does an isolated quantum system evolve toward thermal equilibrium from a far-from-equilibrium initial state—for instance, one prepared by a global quantum quench? This question has attracted significant interest in recent years. The Eigenstate Thermalization Hypothesis (ETH) provides a compelling framework for addressing it, with far-reaching implications from high-energy physics to condensed matter. This primer offers a pedagogical introduction to quantum equilibrium and ETH for a broad physics audience, particularly high-energy physicists seeking a comprehensive understanding of these topics. We assume only basic knowledge of quantum mechanics and statistical mechanics.
\end{abstract}

\vspace{2cm}
\begin{center}
 {\it Based on the review talk presented at the "Workshop on Dynamics and Scrambling of Quantum Information," December 2024, IPM.}
\end{center}

\end{titlepage}

\tableofcontents

\section{Introduction}
\label{sec:intro}

Thermalization of macroscopic systems is one of the most familiar phenomena in everyday experience. Yet thermal behavior can also emerge in isolated quantum systems. Although this problem has been studied since the early days of quantum mechanics \cite{Von}, its precise mechanisms and regimes of validity remain active research topics.

The conceptual difficulty is that closed-system time evolution is unitary and therefore reversible: it preserves purity and all fine-grained information. Consequently, a global pure state never becomes the mixed canonical state $\rho_{\rm th}=e^{-\beta H}/Z(\beta)$. The apparent conflict with statistical mechanics is resolved only when we focus on coarse observables or finite subsystems. Their expectation values---or reduced density matrices---can become indistinguishable from thermal predictions, even though the complete state retains all initial information.

One might therefore expect that unitary quantum mechanics forbids the relaxation of an initial state into an equilibrated one---that true thermal equilibrium should never be reached. Yet thermalization is consistently observed in experiments, from cold atoms to trapped ions~\cite{Kinoshita:2006, Hofferberth:2007, Trotzky:2012, Smith:2016, Zhang:2017, Neyenhuis:2017} (see also reviews~\cite{Yukalov:2011, Polkovnikov:2011, Eisert:2015, Ueda:2018, Altman:2015, Borgonovi:2016}). The resolution is operational: under suitable Hamiltonian and initial-state conditions, local observables and finite-subsystem reduced states may approach thermal predictions in the thermodynamic sense \cite{Dymarsky2018}. The global pure state, however, remains pure under exact unitary evolution.

As we will elaborate, this challenge can be understood through quantum chaos and the structure of energy eigenstates. The Eigenstate Thermalization Hypothesis (ETH) offers a powerful framework for explaining thermalization in isolated quantum systems~\cite{Deutsch:1991, Srednicki:1994mfb, Serdnicki:1999}. In its strong form, ETH states that, within a fixed symmetry sector, bulk eigenstates at finite energy density become locally indistinguishable from the corresponding equilibrium ensemble in the thermodynamic limit. Thermalization after a quench additionally requires an initial state with a sufficiently narrow energy-density distribution and adequate support in that sector. Spectral edges and exceptional nonthermal eigenstates require separate treatment.

The main objective of this primer is to understand 
thermalization in isolated quantum systems. The relevant
information is encoded in the spectrum, eigenstates,
observables, and initial-state weights. After symmetry 
resolution, many quantum-chaotic Hamiltonians exhibit 
random-matrix spectral correlations
\cite{Livan:2017,Bohigas:1983er} and often satisfy ETH for local observables. However, neither property automatically implies the other.

This primer introduces the minimal set of concepts needed to follow the logic of quantum thermalization, with ETH at its core. Its scope is deliberately elementary: we aim to prepare readers for the research literature rather than to provide an exhaustive review. We assume only an undergraduate course in quantum mechanics and some familiarity with statistical mechanics. For broader and more advanced treatments, see Refs.~\cite{DAlessio:2015qtq, Deutsch:2018, Mori:2017qhg} in addition to the original ETH papers~\cite{Deutsch:1991, Srednicki:1994mfb, Serdnicki:1999}. The lecture notes in Ref.~\cite{Pappalardi} offer another useful entry point.

The paper is organized as follows. Section~\ref{sec:equ} defines quantum equilibration using the diagonal ensemble. Section~\ref{sec:thermal} explains when an equilibrated observable is also thermal. Section~\ref{sec:ETH} introduces quantum-chaos diagnostics and the ETH ansatz, illustrated with a finite-size Ising chain. Section~\ref{sec:weak} distinguishes dynamically strong and weak thermalization and describes the restricted family of initial states used in the example. Section~\ref{sec:high} discusses mechanisms of ETH violation and outlines high-energy applications. Section~\ref{sec:outlook} concludes. Appendix~\ref{app:level} covers level spacing and the unfolding-free gap-ratio statistic. Appendices~\ref{app:mathematica} and~\ref{app:julia} provide Mathematica and Julia implementations of the numerical procedures.

%%%%%%%%%%%%%%%%%%%%%%%%%%%%%%%%%%%%%%%%%

\section{Quantum Equilibrium}
\label{sec:equ}

Consider a closed quantum system with a local, time-independent Hamiltonian $H$. Its eigenvalues and eigenstates are denoted by $E_n$ and $\ket{E_n}$, where $n=1,\ldots,\mathcal{D}$ and $\mathcal{D}$ is the Hilbert-space dimension. Whenever exact symmetries are present, all spectral statements below are understood within a fixed irreducible symmetry sector. We assume non-degenerate energies in that sector only where this simplifies the time-average formulas; this is not an assumption of ETH itself.

Initially, we prepare the system in a (non-equilibrium) state described by a density matrix $\rho_0$, which may be pure or mixed. Since the system is isolated, its time evolution under $H$ is unitary:
\be
\rho(t)=e^{-iHt}\rho_0\,e^{iHt}.
\ee

We are primarily interested in the late-time behavior of the expectation value of a typical observable $O$:
\be\label{Ot}
\langle O(t)\rangle = \Tr\left(e^{-iHt}\rho_0\,e^{iHt}O\right).
\ee
For a generic finite system, this expectation value is quasiperiodic and need not converge pointwise. Nevertheless, it can remain close to an equilibrium value with small fluctuations for most times after relaxation. Special cases—such as a stationary density matrix or an observable that commutes with $H$—are of course exactly time-independent.

The central question is therefore: for which observables, initial states, and time windows does a suitable equilibrium ensemble describe the system? Because the exact dynamics is unitary, a finite system does not literally converge to a stationary global density matrix; equilibration must instead be formulated operationally.

The first concept to clarify is what equilibrium means at the quantum level. Following von Neumann's original perspective \cite{Von}, one studies physical observables and finite-subsystem reduced states rather than demanding that the full density matrix become stationary. An observable equilibrates if its expectation value remains close to a stationary value for most sufficiently late times. This behavior is conditional on the Hamiltonian, observable, and initial state; it is not guaranteed for every closed system.

To proceed, we expand the initial density matrix
in the energy eigenstate basis:
\be
\rho_0=\sum_{n,m}\rho_{nm}\ketbra{E_n}{E_m},
\ee
where $\rho_{nn}=\abs{c_n}^2$ for a pure state. Equation \eqref{Ot} then becomes\footnote{The simple fluctuation formula below uses the stronger non-resonance, or non-degenerate-gap, condition: within the selected symmetry sector, equality $E_n-E_m=E_l-E_k$ for nonzero gaps implies $n=l$ and $m=k$. Degeneracies require corresponding projectors and modified bookkeeping, but do not preclude equilibration or ETH.}
\be\label{EOt}
\langle O(t)\rangle = \sum_{n=1}^{\mathcal{D}}\rho_{nn}O_{nn} + \sum_{n\neq m=1}^{\mathcal{D}}\rho_{nm}O_{mn}e^{i(E_n-E_m)t},
\qquad \sum_{n=1}^{\mathcal{D}}\rho_{nn}=1,
\ee
where $O_{mn}=\bra{E_m}O\ket{E_n}$ are the matrix elements of $O$ in the energy basis. The time dependence resides entirely in the off-diagonal terms. Dephasing makes their infinite-time average vanish under the assumptions just stated; equilibration additionally requires that their residual sum be small for most late times, not that it vanish pointwise. The corresponding infinite-time average is\footnote{In our notation, the overline denotes an infinite time average.}
\be\label{Eq-va}
\overline{\langle O(t)\rangle} = \lim_{T\to\infty}\frac{1}{T}\int_0^T dt\,\langle O(t)\rangle
= \sum_{n=1}^{\mathcal{D}}\rho_{nn}O_{nn}
= \Tr(\rho_{\rm DE}O),
\ee
where $\rho_{\rm DE}$ is the density matrix of the diagonal ensemble:
\be
\rho_{\rm DE}=\sum_{n=1}^{\mathcal{D}}\rho_{nn}\ketbra{E_n}{E_n}.
\ee
Thus the long-time equilibrium state is given by the diagonal ensemble:
\be
\bar{\rho}=\lim_{T\to\infty}\frac{1}{T}\int_0^T dt\,e^{-iHt}\rho_0\,e^{iHt}=\rho_{\rm DE}.
\ee

Although the system continues to evolve unitarily, a local observable may appear equilibrated to the diagonal ensemble for most sufficiently late times. Mathematically, after an initial relaxation we require
\be
\abs{\langle O(t)\rangle - \Tr(\rho_{\rm DE}O)} \ll 1.
\ee
Even when this pointwise condition is not satisfied, the time-averaged squared fluctuation may be small:
\be
\overline{\abs{\langle O(t)\rangle - \Tr(\rho_{\rm DE}O)}^2} \ll 1.
\ee
To find an upper bound, we note from \eqref{EOt} and \eqref{Eq-va} that
\be
\overline{\abs{\langle O(t)\rangle - \Tr(\rho_{\rm DE}O)}^2}
= \sum_{n\neq m=1}^{\mathcal{D}}\abs{\rho_{nm}}^2\abs{O_{mn}}^2.
\ee
An immediate bound is therefore
\be\label{bound1}
\overline{\abs{\langle O(t)\rangle - \Tr(\rho_{\rm DE}O)}^2}
\leq \max_{n\neq m}\abs{O_{nm}}^2.
\ee
Thus small off-diagonal matrix elements can suppress late-time fluctuations, provided the initial-state weights populate a sufficiently large portion of the relevant energy shell. This statement concerns fluctuation magnitude; it does not determine the relaxation rate or render the result independent of the initial state.

A more practical bound follows from the inequality $(\rho_{nn}-\rho_{mm})^2\geq 0$, which gives
\be
\overline{\abs{\langle O(t)\rangle - \Tr(\rho_{\rm DE}O)}^2}
\leq \sum_{n\neq m=1}^{\mathcal{D}}\frac{1}{2}(\rho_{nn}^2+\rho_{mm}^2)\abs{O_{nm}}^2
\leq \sum_{n=1}^{\mathcal{D}}\rho_{nn}^2(OO^\dagger)_{nn}.
\ee
To proceed, we recall that for any operator $O$, the spectral norm (or operator norm) is defined as
\be\label{spec-norm-def}
\norm{O}_\infty = \sup_{\ket{\psi}\neq 0} \frac{\norm{O\ket{\psi}}}{\norm{\ket{\psi}}}
= \sup_{\norm{\ket{\psi}}=1} \sqrt{\bra{\psi}O^\dagger O\ket{\psi}}.
\ee
For a Hermitian observable $O=O^\dagger$, this reduces to the largest absolute eigenvalue of $O$:
\be
\norm{O}_\infty = \max_i \abs{\lambda_i},
\ee
where $\lambda_i$ are the eigenvalues of $O$. This norm provides a state-independent upper bound on the magnitude of expectation values: for any state $\ket{\psi}$, $\abs{\bra{\psi}O\ket{\psi}} \leq \norm{O}_\infty$.

For a Hermitian observable, $(OO^\dagger)_{nn} = \bra{E_n}O^2\ket{E_n}$. Since the expectation value of $O^2$ in any state cannot exceed the square of its largest eigenvalue, we have
\be
\bra{E_n}O^2\ket{E_n} \leq \norm{O}_\infty^2.
\ee
Equivalently, by completeness, $\sum_m \abs{O_{nm}}^2 = \bra{E_n}O^2\ket{E_n} \leq \norm{O}_\infty^2$. Since $\abs{O_{nn}}^2 \geq 0$, it follows that the off-diagonal elements satisfy
\be\label{off-O}
\sum_{m(\neq n)=1}^{\mathcal{D}} \abs{O_{nm}}^2 \leq \norm{O}_\infty^2.
\ee
This inequality captures the key physical content: the spectral norm controls the total weight of the off-diagonal matrix elements, which are precisely the terms responsible for time-dependent fluctuations.

Applying this bound to the fluctuation formula, we obtain
\be\label{bound2}
\overline{\abs{\langle O(t)\rangle - \Tr(\rho_{\rm DE}O)}^2}
\leq \sum_{n=1}^{\mathcal{D}}\rho_{nn}^2 \norm{O}_\infty^2
= \frac{\norm{O}_\infty^2}{\xi},
\ee
where $\xi$ is the effective dimension—equivalently, the participation ratio of the initial state in the energy basis \cite{Short:2011pvc}:
\be
\xi^{-1} = \sum_{n=1}^{\mathcal{D}} \rho_{nn}^2 = \Tr(\rho_{\rm DE}^2).
\ee
The effective dimension measures the number of energy eigenstates that contribute significantly to the initial state. It satisfies $1 \leq \xi \leq \mathcal{D}$: $\xi=1$ for a single energy eigenstate, and $\xi=\mathcal{D}$ for uniform weight over the entire spectrum. The bound \eqref{bound2} therefore shows that the long-time-averaged fluctuation of any bounded observable is suppressed by the effective dimension. A large effective dimension—meaning that the initial state populates many energy eigenstates—guarantees small fluctuations around the diagonal-ensemble prediction. For a state with small effective dimension the bound is weak or trivial and supplies no conclusion. In particular, a single energy eigenstate is exactly stationary, so $\xi=1$ must not be interpreted as an absence of equilibration.

It is important to emphasize what this bound does and does not imply. First, it is a statement about the time-averaged fluctuations, not about the instantaneous behavior at any specific time. An observable could exhibit large fluctuations at isolated times while still satisfying the bound on average. Second, the bound does not determine the relaxation time scale; it only constrains the long-time behavior. Third, the bound is universal in the sense that it applies to any bounded observable and any initial state, but it is not tight—the actual fluctuations may be much smaller than the bound suggests. Fourth, and most crucially, the bound concerns equilibration to the diagonal ensemble, not thermalization. Even if the fluctuations are small, the equilibrium value $\Tr(\rho_{\rm DE}O)$ may still differ significantly from the thermal prediction. Additional structure, such as the smoothness of diagonal matrix elements encoded in ETH, is required to ensure that the equilibrium value is thermal.

If a closed quantum system equilibrates, Eq.~\eqref{Eq-va} gives its stationary observable values. Formally, when every diagonal weight is nonzero, the diagonal ensemble can be written as
\be
\rho_{\rm DE}=\exp\left(-\sum_{n=1}^{\mathcal{D}}\zeta_n P_n\right),
\ee
where $P_n=\ket{E_n}\bra{E_n}$ and $\zeta_n=-\ln\rho_{nn}$; zero weights are obtained in the limit $\zeta_n\to+\infty$. This is a tautological exponential representation using all spectral projectors, not a predictive generalized Gibbs ensemble. A conventional GGE is useful only when a physically motivated, complete set of local or quasilocal conserved charges provides a compressed description. The exact state remains pure whenever the initial state is pure, even if selected observables equilibrate.

Equilibration to the diagonal ensemble can occur when the relevant off-diagonal matrix elements and initial-state coherences produce small late-time fluctuations, or when a bounded observable is combined with a large effective dimension. These are sufficient mechanisms, not universal guarantees. Additional structure is required for the equilibrium value to be thermal; that is the subject of the next section.

%%%%%%%%%%%%%%%%%%%%%%%%%%%%%%%%%%%%%

\section{Quantum Thermal Equilibrium}
\label{sec:thermal}

For a large quantum many-body system, equilibrium values of local observables can often be described by a few thermodynamic parameters such as energy, temperature, and particle number. An observable is thermalized when, after relaxation, it remains close for most times to the prediction of an appropriate microcanonical or canonical ensemble. This does not require the global diagonal density matrix to be close to a thermal density matrix in trace norm. Rather, the ensembles must be indistinguishable for a specified class of local or few-body observables, or equivalently on fixed finite subsystems in the thermodynamic limit. The framework applies to both pure and mixed initial states, although their populations and coherences can lead to different equilibration behavior.

To make this statement more precise, suppose the system is prepared in an initial state described by $\rho_0$. The average energy of the state is then
\be
E_0 = \Tr(\rho_0 H) = \sum_{n=1}^{\mathcal{D}}\rho_{nn}E_n.
\ee
Choose a microcanonical shell $[E_0-\Delta E, E_0+\Delta E]$ that contains many levels while remaining narrow in energy density. The associated density matrix is
\be
\rho_{\rm MC} = \frac{1}{d_{\rm MC}}\sum_{E_\alpha\in[E_0-\Delta E, E_0+\Delta E]}\ketbra{E_\alpha}{E_\alpha},
\ee
where $d_{\rm MC}$ is the number of levels in the shell. A symmetric discrete shell need not have mean energy exactly $E_0$, but its mismatch is negligible on the energy-density scale for a suitable thermodynamic sequence. If $\varrho(E)$ is the density of states and varies slowly across the window, then $d_{\rm MC}\simeq 2\Delta E\,\varrho(E_0)$ and $\log d_{\rm MC}=S(E_0)+\mathcal{O}(N)$, including a subextensive window-dependent term.

Alternatively, we may describe the system using a canonical ensemble with density matrix
\be\label{rho-th}
\rho_{\rm th} = \frac{e^{-\beta H}}{Z(\beta)},\qquad Z(\beta)=\Tr(e^{-\beta H}),
\ee
where the inverse temperature $\beta$ is fixed by the condition $\Tr(\rho_{\rm th}H)=E_0$.

For a given operator, the ensemble expectation values are
\begin{align}\label{MC-CA}
\langle O\rangle_{\rm MC} &= \Tr(\rho_{\rm MC}O) = \frac{1}{d_{\rm MC}}\sum_{E_n\in[E_0-\Delta E, E_0+\Delta E]} O_{nn},
\nonumber\\
\langle O\rangle_{\rm th} &= \Tr(\rho_{\rm th}O) = \frac{1}{Z(\beta)}\sum_{n=1}^{\mathcal{D}}e^{-\beta E_n}O_{nn},
\end{align}
which may be contrasted with Eq.~\eqref{Eq-va}. Thermalization requires that, for every observable $O$ in the stated local or few-body class,
\be\label{Oth}
\Tr[(\rho_{\rm DE}-\rho_{\rm MC})O]\longrightarrow 0,
\qquad\text{or}\qquad
\Tr[(\rho_{\rm DE}-\rho_{\rm th})O]\longrightarrow 0,
\ee
in the thermodynamic limit, with the usual finite-size qualifications. Matching the mean energy alone is necessary but not sufficient for thermalization. Equation~\eqref{Oth} is often associated with quantum ergodicity, but quantum-chaos diagnostics do not by themselves imply it \cite{Deutsch:1991}.

Quantum integrability has no single definition that covers every many-body model. It should not be identified merely with one or several conservation laws: non-integrable systems also conserve energy and may conserve particle number. In many integrable lattice models, there exists an extensive, sufficiently complete family of mutually commuting local or quasilocal charges---for example, those generated by a transfer matrix in a Bethe-integrable model. After exact symmetries are resolved, Poissonian statistics are a common diagnostic rather than an operational definition; they also occur in other nonergodic settings. These constraints can prevent relaxation to an ordinary microcanonical or canonical ensemble. For appropriate initial states, local observables may instead equilibrate to a generalized Gibbs ensemble (GGE) \cite{Rigol:2007}:
\be
\rho_{\rm GGE} = \frac{1}{Z_{\rm GGE}}\exp\left(-\sum_i \lambda_i I_i\right),
\ee
where the physically relevant local and quasilocal charges $I_i$ must be sufficiently complete, and the multipliers $\lambda_i$ are fixed by the initial state.

For further use, we note that for systems with an exponentially large density of states, the canonical factor $e^{-\beta H}$ causes the expectation value to peak at a specific energy, closely resembling a microcanonical ensemble. To elaborate, for an arbitrary smooth function $f(E_n)$, the canonical sum can be replaced by an integral:
\be
\langle f\rangle_{\rm th} = \frac{1}{Z(\beta)}\sum_{n=1}^{\mathcal{D}}e^{-\beta E_n}f(E_n)
\simeq \frac{1}{Z(\beta)}\int_{E_{\min}}^{E_{\max}} dE\,e^{S(E)-\beta E}f(E).
\ee
For a system of size $N$ with extensive entropy, this integral may be evaluated by the saddle-point method. For a smooth intensive function, $\langle f\rangle_{\rm th}=f(E_\beta)+\mathcal{O}(1)$ as $N\to\infty$, with corrections controlled by derivatives of $f$ and the canonical energy variance (often algebraic in $N$, rather than exponentially small in $\mathcal{D}$). The saddle satisfies $S'(E_\beta)=\beta$. The inverse temperature is chosen exactly from $\langle H\rangle_\beta=E_0$; the saddle energy and this mean agree up to ordinary finite-size corrections.

In order to see thermalization, two logically distinct ingredients are therefore needed. First, the time-dependent off-diagonal contribution must be small for most late times, so that the observable equilibrates to its diagonal-ensemble value. Second, Eq.~\eqref{Oth} must hold for the relevant observables, so that this equilibrium value is thermal.

These conditions raise two apparent conceptual problems. First, the many-body mean level spacing is exponentially small, so the Heisenberg time---the inverse mean level spacing, up to conventional factors of $2\pi$---is exponentially large. Local observables need not wait until this time to relax. Their relaxation time is instead controlled by microscopic interaction scales, transport or hydrodynamic modes, and the frequency dependence of the relevant off-diagonal matrix elements. It may therefore grow algebraically with system size, or remain of microscopic order, even though exact recurrences and complete spectral resolution occur only on much longer timescales. 
%There is no universal ``short'' thermalization time; it depends on the Hamiltonian, observable, conserved quantities, and initial state.

The second issue concerns Eq.~\eqref{Oth}, which appears contradictory. Equations \eqref{Eq-va} and \eqref{MC-CA} show that computing the expectation value of an operator in different ensembles involves evaluating the diagonal matrix elements $O_{nn}$ in the energy basis and summing them with a suitable weight. For the diagonal ensemble, this weight is explicitly determined by the probabilities $\rho_{nn}$ that the initial state occupies energy eigenstate $\ket{E_n}$, which remain constant over time. In other ensembles, however, information about the initial state is encoded only indirectly, in a coarse-grained way through the energy or temperature. Thus Eq.~\eqref{Oth} cannot hold universally. Nevertheless, there are approaches to resolve these apparent conceptual inconsistencies.

A useful way to formulate the required structure is to express the matrix elements of a local observable in terms of
\be\label{barE}
\bar{E} = \frac{E_n+E_m}{2},\qquad \omega = E_n-E_m.
\ee
For an initial state with mean energy $E_0$, define its energy variance as
\be
\sigma_E^2 = \Tr[\rho_0(H-E_0)^2] = \sum_n \rho_{nn}(E_n-E_0)^2.
\ee
The state is narrow in energy density when $\sigma_E/N\to 0$. In particular, its probability is concentrated inside every fixed energy-density window: for each fixed $\epsilon>0$,
\be
\sum_{\abs{E_n-E_0}>\epsilon N}\rho_{nn}\longrightarrow 0
\qquad (N\to\infty).
\ee
The variance condition implies this concentration by Chebyshev's inequality. The criterion is invariant under shifting the zero of energy. Product states of finite-range local Hamiltonians commonly have $\sigma_E=\mathcal{O}(\sqrt{N})$ and therefore satisfy it. Smooth diagonal matrix elements then give
\be
\sum_n \rho_{nn}O_{nn} \approx \widetilde O(E_0)\sum_n \rho_{nn} = \widetilde O(E_0)
\ee
up to corrections controlled by the energy-density width. The exponential suppression of typical off-diagonal elements does not follow from smoothness alone. As we will see, in chaotic many-body systems, it is an additional statistical statement, motivated by random-matrix behavior and encoded explicitly in the ETH ansatz of the next section. The smooth frequency envelope controls dephasing and relaxation, whereas the exponentially small level spacing controls the much later Heisenberg and recurrence scales.

To conclude, thermalization requires two distinct ingredients: equilibration, meaning that the off-diagonal contribution is small for most sufficiently late times; and agreement of the diagonal ensemble with the thermal ensemble for the chosen observable class. Smooth diagonal matrix elements over the narrow energy-density distribution of the initial state support the second ingredient. The entropic scaling and frequency dependence of off-diagonal elements are additional statistical statements supplied by ETH, not consequences of diagonal smoothness. The next section makes these statements more precise.

It is crucial to emphasize, however, that throughout this primer, ``thermalization'' always refers to the behavior of \emph{local} (or few-body) observables and to the reduced density matrices of finite subsystems. The global pure state of an isolated quantum system never thermalizes; it remains pure under unitary evolution. What becomes thermal are the expectation values of local operators and the reduced states of small subsystems.

%%%%%%%%%%%%%%%%%%%%%%%%%%%%%%%%%%%%
%%%%%%%%%%%%%%%%%%%%%%%%%%%%%%%%%%5
\section{ETH and Quantum Chaos}
\label{sec:ETH}

ETH is a statistical ansatz for matrix elements of local observables in the energy basis. Within its domain of validity, it explains why individual bulk eigenstates exhibit thermal local properties and why suitable superpositions equilibrate to the corresponding ensemble. This section first distinguishes ETH from complementary diagnostics of quantum chaos and then states the ansatz precisely.

\subsection{Quantum Chaos and Its Diagnostics}

Quantum chaos has no single universally sufficient definition for finite many-body systems. Unlike classical chaos, which admits a clear characterization through exponential sensitivity to initial conditions, the quantum realm lacks a direct analogue of trajectories in phase space. Operationally, one therefore resolves every exact symmetry and combines several complementary diagnostics: random-matrix spectral correlations (including level spacing, gap ratio, and the spectral form factor), eigenvector and observable-matrix statistics, and dynamical probes such as operator spreading and out-of-time-ordered correlators (OTOCs). Each probes different aspects of the system, and none alone proves ETH or thermalization. The classical notion of sensitive dependence provides useful semiclassical motivation, but classical chaos, ergodicity, and thermalization are not logically equivalent in full generality. A system can be classically chaotic yet fail to thermalize quantum mechanically due to, for example, integrability-breaking perturbations that are too weak, or the presence of additional conserved quantities that constrain the dynamics.

Classical chaotic behavior is characterized by the sensitivity of trajectories in phase space to initial conditions: two initially nearby trajectories separate exponentially fast, quantified by the Lyapunov exponent. Denoting the collective coordinates of phase space by $Y$, a small change in initial conditions, $Y(0)\to Y(0)+\delta Y(0)$, leads to
\be\label{CC}
\norm{\delta Y(t)} \sim \norm{\delta Y(0)} e^{\lambda_{\rm cl}t},
\ee
where $\lambda_{\rm cl}>0$ is the largest classical Lyapunov exponent. In typical stable regular motion, by contrast, nearby trajectories remain bounded or separate at most polynomially \cite{Book:2009}. This exponential sensitivity is the hallmark of classical chaos, but it is not a converse definition: a hyperbolic fixed point can produce local exponential separation without global chaotic dynamics. Moreover, the presence of a positive Lyapunov exponent in the classical limit does not guarantee that the corresponding quantum system will exhibit thermalization, as quantum effects such as interference and tunneling can suppress or modify chaotic behavior.

Semiclassically, sensitivity to initial conditions relates to a Poisson bracket, $\partial Y(t)/\partial Y(0) \sim \{Y(t), Y(0)\}$, which after quantization motivates the positive squared-commutator diagnostic
\be
C(t) = -\left\langle [O(t), Q(0)]^2 \right\rangle_\beta \geq 0
\ee
for Hermitian operators $O$ and $Q$. Expanding the commutator expresses $C(t)$ in terms of ordinary correlators and an out-of-time-ordered correlator (OTOC), such as
\be
F(t) = \left\langle O(t) Q(0) O(t) Q(0) \right\rangle_\beta.
\ee
Here $\langle\cdots\rangle_\beta = \Tr(e^{-\beta H}\cdots)/Z(\beta)$ denotes the canonical thermal average. The OTOC measures the degree to which the operators $O(t)$ and $Q(0)$ fail to commute, providing a diagnostic of information scrambling and operator growth. The displayed $F(t)$ is a common unregulated OTOC. The Maldacena--Shenker--Stanford theorem applies more specifically to an appropriately normalized, thermally regulated OTOC of bounded operators, for example
\be
F_{\rm reg}(t)=\Tr\!\left[yO(t)yQ(0)yO(t)yQ(0)\right],
\qquad y^4=\frac{e^{-\beta H}}{Z(\beta)}.
\ee
In a controlled semiclassical or large-parameter regime at positive inverse temperature $\beta>0$, suppose that the connected deviation of this correlator has a small exponential contribution $\varepsilon e^{\lambda_{\rm L}t}$ in a window $t_d\ll t\ll t_*$ between dissipation and scrambling. Under the required thermal analyticity, boundedness, and factorization assumptions, its exponent obeys the MSS bound,
\be
\lambda_{\rm L} \leq \frac{2\pi}{\beta},
\ee
in units $\hbar = k_{\rm B} = 1$ \cite{Maldacena:2015waa}. The result is a bound within this domain of assumptions, not a bound on arbitrary unregulated OTOCs or on every measure of operator growth. Semiclassical Einstein black holes and certain strongly coupled theories saturate it in the appropriate regime.

It is crucial to recognize that exponential growth of OTOCs is not a definitive indicator of chaos. Such growth can also occur in certain non-chaotic systems, such as the inverted harmonic oscillator, which exhibits exponential sensitivity despite being integrable \cite{Hashimoto:2020xfr}. Similarly, OTOC growth can be modified by the presence of conservation laws, symmetries, or finite-size effects. This serves as a reminder that quantum chaos is a subtle concept requiring careful diagnostics and that no single observable provides a complete characterization.

Spectral statistics provide a complementary and historically important diagnostic of quantum chaos \cite{Haake:2010,Stockmann:1999}. Before computing them, the Hamiltonian must be block-diagonalized into irreducible symmetry sectors; mixing independent sectors can spuriously produce Poisson-like statistics and invalidates comparison with a single random-matrix class. Within a fixed sector, order the eigenvalues as $E_{n+1} > E_n$ and define the spacings $s_n = E_{n+1} - E_n$. After unfolding (Appendix~\ref{app:level}), integrable spectra are commonly compared with Poisson statistics, whereas spectra with random-matrix correlations are compared with the Wigner--Dyson class selected by the antiunitary symmetries acting within that sector \cite{Wigner:1955, Wigner:1957, Wigner:1958,
Mehta:2004}. Common nearest-neighbor benchmarks are the Poisson law and the $2\times2$ Wigner surmises,
\bea
\text{Poisson:} &\quad P(s) = e^{-s},\\
\text{Wigner surmise:} &\quad P(s) = A s^\delta e^{-Bs^2},\qquad \delta = 1,2,4,
\eea
where the constants $A$ and $B$ are fixed by the normalization conditions
\be
\int_0^\infty P(s)\,ds = 1,\qquad \int_0^\infty s P(s)\,ds = 1,
\ee
which enforce unit total probability and unit mean level spacing, respectively. Poisson statistics have no level repulsion and commonly occur in integrable systems after all symmetries are resolved; the Wigner surmises exhibit the level repulsion characteristic of their random-matrix ensembles. The exponent $\delta$ is fixed by the antiunitary symmetry class of the resolved block: $\delta=1$ when an antiunitary operator $T$ maps the block to itself and satisfies $T^2=+1$ (Gaussian orthogonal ensemble, GOE); $\delta=2$ when no such antiunitary symmetry acts within the block (Gaussian unitary ensemble, GUE); and $\delta=4$ when $T^2=-1$ (Gaussian symplectic ensemble, GSE), with Kramers pairs treated consistently \cite{Mehta:2004}. Thus ordinary ``time-reversal symmetry'' and the presence of half-integer spin are not, by themselves, a complete classification. These distributions are benchmarks for particular spectral correlations, not a scalar scale on which being ``closer'' to a Wigner curve means that a model is more chaotic. Level statistics also do not determine eigenstate structure or observable dynamics.

In modern many-body physics, the \emph{gap ratio} statistic is often preferred over level spacing because it does not require spectral unfolding. The gap ratio is defined from consecutive level spacings and is insensitive to the local density of states, making it particularly useful for systems with small Hilbert-space dimensions or complicated density of states where unfolding is unreliable. This is discussed in detail in Appendix~\ref{app:level} \cite{Atas:2013, Oganesyan:2007}. The gap ratio has become the standard diagnostic for quantum chaos in many-body physics due to its simplicity and robustness.

Quantum-chaotic systems with Wigner--Dyson level statistics often also satisfy ETH for local observables, and random-matrix ideas motivate the statistical part of the ETH ansatz \cite{Deutsch:1991, Srednicki:1994mfb, Serdnicki:1999}. However, the implication is not a theorem: level statistics probe eigenvalues, whereas ETH constrains eigenvectors and operator matrix elements. A system can exhibit Wigner--Dyson statistics yet fail ETH for a specified observable. Conversely, finite-size crossovers or additional dynamical structure can produce deviations from the random-matrix benchmark without by themselves deciding whether ETH holds. Unresolved exact symmetries are not such a counterexample; they make the spectral test improperly posed and must first be resolved. Spectral statistics, eigenstate properties, and dynamical probes should therefore be examined independently and then compared.

\subsection{The ETH Ansatz}

Consider a quantum many-body system of size $N$, with Hilbert-space dimension $\mathcal{D}$, described by a generic local Hamiltonian $H$. Work within a fixed irreducible symmetry sector and choose an orthonormal energy basis, resolving any protected degeneracy consistently. For a local or few-body observable, the ETH ansatz reads \cite{Deutsch:1991, Srednicki:1994mfb, Serdnicki:1999, DAlessio:2015qtq}
\be\label{ETH}
O_{nm} = \bra{E_n}O\ket{E_m}
= \tilde{O}(\bar{E})\delta_{nm} + e^{-S(\bar{E})/2} f_O(\bar{E},\omega) R_{nm}.
\ee
Here $\bar{E}$ and $\omega$ are defined in Eq.~\eqref{barE}, $S(\bar{E})=\mathcal{O}(N)$ is the thermodynamic entropy, and $\tilde{O}(\bar{E})$ and $f_O(\bar{E},\omega)$ are smooth envelopes.

The structure of the ansatz is worth examining in detail. The diagonal term, $\tilde{O}(\bar{E})\delta_{nm}$, represents the smooth, thermal expectation value of the observable at energy $\bar{E}$. It depends only on the average energy and is responsible for the thermal properties of individual eigenstates. The off-diagonal term consists of three factors. The exponential factor $e^{-S(\bar{E})/2}$ ensures that off-diagonal matrix elements are exponentially small in the system size, scaling as $e^{-O(N)}$. This suppression is essential for equilibration: it guarantees that the time-dependent fluctuations of local observables vanish in the thermodynamic limit. The smooth envelope $f_O(\bar{E},\omega)$ encodes the frequency dependence of these fluctuations, which controls the dynamics of relaxation, dephasing, and transport. Finally, the random variable $R_{nm}$ captures the universal statistical fluctuations of the matrix elements.

The quantities $R_{nm}$ are locally normalized pseudorandom residuals rather than entries that must be drawn independently from a literal random matrix. Their smooth-window averages are conventionally chosen as $\overline{R_{nm}}=0$ and $\overline{\abs{R_{nm}}^2}=1$, with Hermiticity imposing $R_{mn}=R_{nm}^*$ after a consistent convention for $f_O$. In an antiunitary class with a real representation the residuals may be taken real, whereas a complex class generally requires complex residuals. Gaussian-like distributions are observed in suitable energy and frequency windows in a number of non-integrable models \cite{Beugeling:2015,Mondaini:2017}, but exact Gaussianity is not part of ETH and is not required by Eq.~\eqref{ETH}. Finite-size effects, proximity to integrability, conservation laws, kinetic constraints, or other structure can all produce non-Gaussian residuals. Many-body-localized systems can exhibit broad or sparse matrix-element distributions \cite{Basko:2006, Serbyn:2013, Huse:2014}, but non-Gaussianity alone neither establishes localization nor proves a failure of thermalization. Distributional tests must therefore be performed locally in $(\bar E,\omega)$ and interpreted together with other diagnostics.

The ETH ansatz is sufficient to achieve thermal equilibrium. Assuming ETH, one can show that the off-diagonal matrix elements are exponentially small and that the resulting thermalized value is consistent with the microcanonical ensemble. These are precisely the two crucial ingredients for thermalization identified in previous sections: small off-diagonal elements ensure equilibration to the diagonal ensemble, while the smoothness of the diagonal elements ensures that the diagonal ensemble agrees with the thermal ensemble. For an initial state supported in a narrow energy-density window, ETH supplies both ingredients. For a pure state with coefficients $c_n$ in the energy basis, the non-degenerate-gap result gives
\be
\overline{\abs{\langle O(t)\rangle - \Tr(\rho_{\rm DE}O)}^2}
= \sum_{n\neq m} \abs{c_n}^2\abs{c_m}^2 e^{-S(\bar{E})} \abs{f_O(\bar{E},\omega)}^2 \abs{R_{nm}}^2
= \mathcal{O}(e^{-S(E_0)}),
\ee
provided the smooth envelope and local moments of $R_{nm}$ remain bounded in the window. Thus temporal fluctuations of a local observable are exponentially suppressed in system size for typical ETH data.

To explore how thermal equilibrium arises, we compute the thermal expectation value $\langle O\rangle_{\rm th}$ using ETH. Inserting the diagonal part of the ansatz into the canonical expectation value yields
\be
\langle O\rangle_{\rm th} = \frac{1}{Z(\beta)}\sum_{n=1}^{\mathcal{D}} e^{-\beta E_n}O_{nn}
= \frac{1}{Z(\beta)}\int dE\,e^{S(E)-\beta E}\tilde{O}(E) + \mathcal{O}(e^{-S(E_\beta)/2}).
\ee
The saddle-point argument of the previous section then gives
\be
\langle O\rangle_{\rm th} = \tilde{O}(E_0) +\mathcal{O}(1) 
+ \mathcal{O}(e^{-S(E_0)/2}).
\ee
The same smoothness argument gives $\langle O\rangle_{\rm MC}\approx \tilde{O}(E_0)$ for a narrow microcanonical shell. It remains to compare this with the diagonal ensemble.

Expand the smooth diagonal envelope about $E_0$ and define the moments of the energy distribution:
\be
\mu_k = \sum_n \rho_{nn}(E_n-E_0)^k,\qquad \mu_1 = 0.
\ee
Then the diagonal ensemble expectation value is
\be\label{Ode}
\Tr(\rho_{\rm DE}O) = \tilde{O}(E_0) + \sum_{k=2}^{\infty}\frac{\mu_k}{k!}\tilde{O}^{(k)}(E_0) +
\mathcal{O}(e^{-S(E_0)/2}).
\ee
The diagonal and thermal ensembles agree when the correction term is small compared with the natural scale of the observable, $O_{\rm sc}$:
\be\label{DE}
\abs{\sum_{k=2}^{\infty}\frac{\mu_k}{k!}\tilde{O}^{(k)}(E_0)} \ll O_{\rm sc}.
\ee
When this condition holds, the correction can be dropped, yielding
\be
\Tr(\rho_{\rm DE}O) \approx \tilde{O}(E_0),
\ee
which fulfills the second condition for thermalization. A standard sufficient condition for this to hold is $\sigma_E = \mathcal{O}(N)$—that is, the energy standard deviation is subextensive, or equivalently $\sigma_E^2=\mathcal{O}(N^2)$—together with concentration of the energy distribution in a narrow energy-density window and smooth thermodynamic scaling of $\tilde{O}$. These conditions are satisfied by broad classes of product states for finite-range local Hamiltonians, for which typically $\sigma_E = 
\mathcal{O}(\sqrt{N})$.

Even in this approximation, the diagonal matrix is not exactly proportional to the identity. To first order in energy about $E_0$, we have
\be\label{ETH0}
O_{nm} \approx \tilde{O}(E_0)\delta_{nm}
+ \tilde{O}'(E_0)\bra{E_n}H - E_0\mathbf{1}\ket{E_m}
+ e^{-S(\bar{E})/2} f_O(\bar{E},\omega) R_{nm}.
\ee

Under the assumptions stated above, ETH therefore implies that for most post-relaxation times,
\be\label{TO}
\bra{\psi(t)}O\ket{\psi(t)} \approx \Tr(\rho_{\rm th}O) + \text{small fluctuations},
\ee
for most times after relaxation and before finite-size recurrences. No ensemble average over Hamiltonians is required. The conclusion depends on the initial energy weights through their narrow energy-density support, as quantified by Eq.~\eqref{DE}, and on the smoothness of $\tilde{O}$.

The standard ETH ansatz applies to local or few-body observables in bulk finite-energy-density windows. Ground states, low-lying excitations, and the extreme upper edge are normally outside this domain. Conventional ETH also generally fails in integrable systems, whose additional conserved charges require a different ensemble description.

The physical content of strong ETH is that individual finite-energy-density eigenstates are locally indistinguishable, in the thermodynamic limit, from the appropriate thermal ensemble for the class of observables under consideration. This does not make a global pure eigenstate equal to a mixed thermal density matrix. Unitary dynamics dephases coherences between populated eigenstates, allowing the local thermal information encoded in their diagonal matrix elements to control late-time observables.

A crucial point is that the static ETH ansatz by itself does not determine every dynamical timescale. The frequency envelope $f_O(\bar{E},\omega)$ contains important dynamical information: its low-frequency structure can encode transport and hydrodynamic modes, while its high-frequency tail constrains short-time behavior. Nevertheless, the width of $f_O$ alone does not fix a unique relaxation rate; the initial spectral weights, phases, conservation laws, and finite-size scales also enter. With a specified normalization and density-of-states convention, $\abs{f_O(\bar E,\omega)}^2$ is related to the microcanonical correlation spectral density. Detailed balance and linear response then give a fluctuation--dissipation relation between appropriate equilibrium correlation and response functions; it is not, in general, precise to say that $f_O$ itself universally ``satisfies'' the fluctuation--dissipation theorem. Predicting a relaxation curve therefore requires the envelope together with the initial state and the relevant transport data.

%%%%%%%%%%%%%%%%%%%%%%%%%%%%%%%%%%%%%%%%%

\subsection{Numerical Evidence}

\begin{figure}[!ht]
	\begin{center}
		\includegraphics[width=0.45\linewidth]{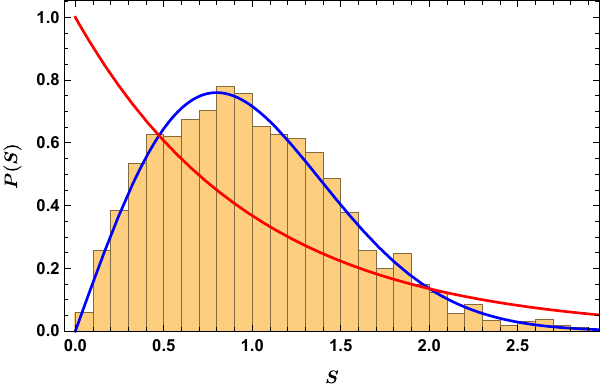}
		\includegraphics[width=0.45\linewidth]{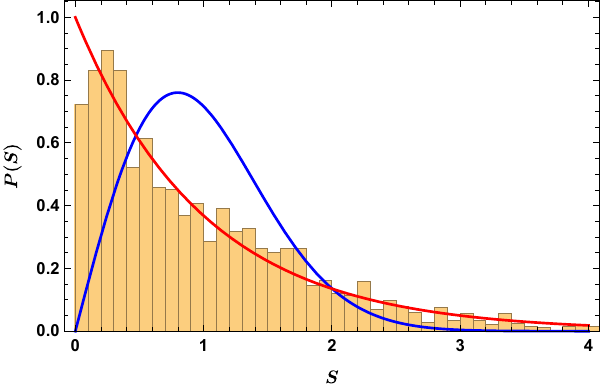}
	\end{center}
	\caption{Level-spacing distributions for Eq.~\eqref{Ising0} at $N=10$ and $g=-1.05$. The mixed-field data (left, $h=0.5$) use the reflection-even sector of dimension 528 and are compared with the GOE Wigner surmise, $P(s)=\frac{\pi}{2}s e^{-\pi s^2/4}$. The integrable data (right, $h=0$) use the joint reflection-even, spin-flip-even sector of dimension 272 and are compared with $P(s)=e^{-s}$. The full Hilbert-space dimension is 1024, and the spectra are unfolded as described in Appendix~\ref{app:level}.}
	\label{fig:level}
\end{figure}

For a concrete finite-size illustration, consider the open spin-$1/2$ Ising chain
\be\label{Ising0}
H = -J\sum_{i=1}^{N-1}\sigma^z_i\sigma^z_{i+1} - \sum_{i=1}^N (g\,\sigma^x_i + h\,\sigma^z_i).
\ee
Here $\sigma^{x,y,z}$ are Pauli matrices 
and $J$, $g$, and $h$ define the model.
By rescaling, we set $J=1$. For generic nonzero 
transverse and longitudinal fields, the model 
is non-integrable; $h=0$ gives the integrable transverse-field Ising chain. Spectral statistics must be computed within fully resolved symmetry sectors. At open boundaries, the Hamiltonian has reflection symmetry; at $h=0$, it also has the global spin-flip symmetry $F=\prod_i\sigma_i^x$. Setting $g=-1.05$, Figure~\ref{fig:level} compares the mixed-field model in the reflection-even sector with the integrable model in the joint reflection-even, spin-flip-even sector. The former is expected to follow GOE statistics because the block has an anti-unitary symmetry squaring to $+1$; the latter is compared with Poisson statistics \cite{Banuls:2011vuw}.

To probe the eigenstate structure entering ETH with an observable whose support
and normalization do not grow with system size, we use, for even $N$,
\be
O_c=\frac{\sigma^x_{N/2}+\sigma^x_{N/2+1}}{\sqrt{2}}.
\ee
This operator is local and reflection even, and its normalized
Hilbert--Schmidt norm satisfies $2^{-N}\Tr(O_c^\dagger O_c)=1$ for every even
$N$. Each matrix-element calculation is performed within a fixed block of the
explicitly resolved global symmetries, so that zeros connecting different
symmetry sectors are not mistaken for ETH suppression.

\begin{figure}[!ht]
	\centering
\includegraphics[width=0.45\linewidth]{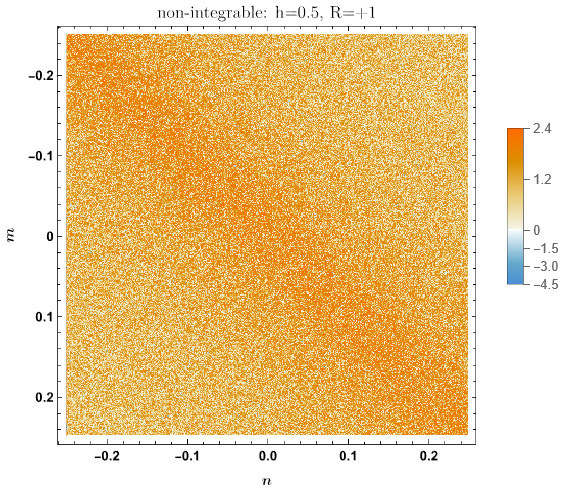}
   \includegraphics[width=0.45\linewidth]{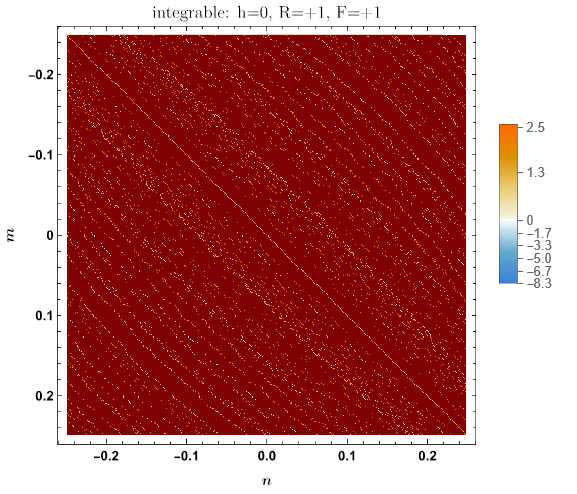}
	\caption{Symmetry-resolved bulk matrix elements of
	$O_c=(\sigma^x_{N/2}+\sigma^x_{N/2+1})/\sqrt{2}$ for the open Ising chain at
	$N=12$, $J=1$, and $g=-1.05$; here
	$O_c=(\sigma^x_6+\sigma^x_7)/\sqrt{2}$. Only eigenstates satisfying
	$\abs{E-E_\infty}\leq3J$ are displayed, where
	$E_\infty=\Tr(H_{\rm sec})/d_{\rm sec}$. Left: the non-integrable $h=0.5$,
	reflection-even ($R=+1$) block, with $d_{\rm sec}=2080$ and
	$d_{\rm win}=884$.
	Right: the integrable $h=0$, reflection-even and spin-flip-even block, with
	$R=F=+1$, $d_{\rm sec}=1056$, and $d_{\rm win}=466$. The common color scale is
	$\log_{10}[\sqrt{d_{\rm win}}\abs{(O_c)_{nm}}]$. The diagonal is masked in
	white; maroon denotes $\abs{(O_c)_{nm}}\leq10^{-12}$. The off-diagonal
	numerical-zero fractions are $0.00\%$ and $96.44\%$, respectively.}
	\label{fig:Ising1}
\end{figure}

Figure~\ref{fig:Ising1} compares the bulk matrix of $O_c$ in the
non-integrable and integrable chains at $N=12$. The diagonal is deliberately
masked, and the two panels share a logarithmic scale for
$\sqrt{d_{\rm win}}\abs{(O_c)_{nm}}$, where $d_{\rm win}$ is the number of
eigenstates retained in the displayed energy window. The non-integrable block
has dense, irregular support at the numerical resolution used here. By
contrast, the integrable block remains highly sparse and exhibits pronounced
banded structure even after reflection and spin-flip symmetries have been
resolved: $96.44\%$ of its displayed off-diagonal entries lie below the
numerical threshold. These are numerical zeros rather than a claim that every
such entry vanishes exactly. The contrast is a finite-size structural
diagnostic; a heat map alone is not a proof of the ETH scaling ansatz.

A quantitative ETH test must retain both arguments of the off-diagonal
envelope. With $\bar E=(E_n+E_m)/2$ and $\omega=E_n-E_m$, define
\be
V_{O_c}(\bar E,\omega)
=\left\langle\abs{(O_c)_{nm}}^2\right\rangle_{\bar E,\omega},
\ee
where the average is taken over unordered pairs in a narrow
$(\bar E,\abs{\omega})$ bin. Equation~\eqref{ETH} predicts
\be
V_{O_c}(\bar E,\omega)
\simeq e^{-S(\bar E)}\abs{f_{O_c}(\bar E,\omega)}^2.
\ee
At finite size, we use the shell population
$D_{\rm mc}=\#\{n:\abs{E_n-E_\infty}\leq\Delta E\}$ as a density-of-states
proxy. Consequently, $D_{\rm mc}V_{O_c}$ is a finite-window proxy for
$e^{S(\bar E)}V_{O_c}$, not an exact identification with the thermodynamic
quantity.

\begin{figure}[!ht]
	\centering
\includegraphics[width=0.45\linewidth]{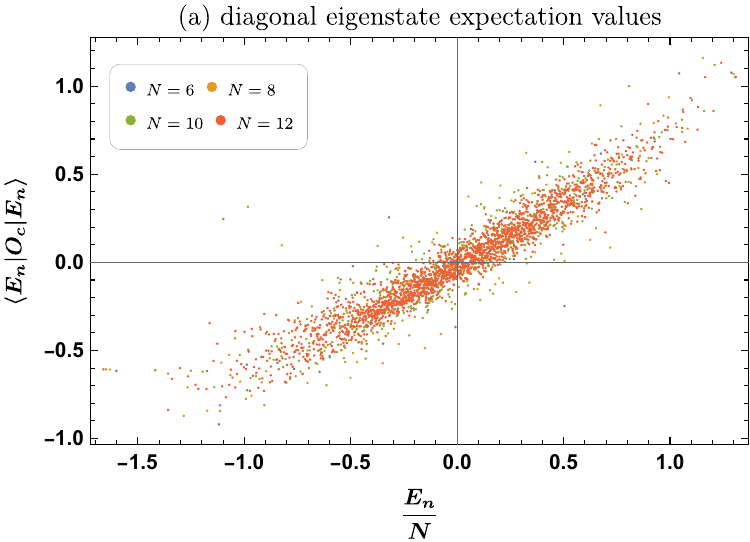}
\includegraphics[width=0.45\linewidth]{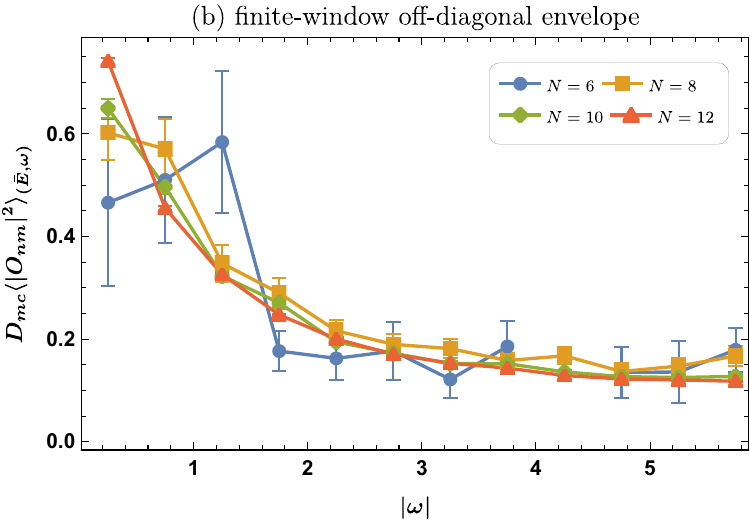}
	\caption{Symmetry-resolved ETH diagnostics for
	$O_c=(\sigma^x_{N/2}+\sigma^x_{N/2+1})/\sqrt{2}$ in the reflection-even
	block ($R=+1$) of the non-integrable chain ($J=1$, $g=-1.05$, $h=0.5$) at
	$N=6,8,10,12$, whose sector dimensions are $36$, $136$, $528$, and $2080$.
	(a) Diagonal eigenstate expectation values versus energy density for the
	full sector spectra. (b) $D_{\rm mc}V_{O_c}(\bar E,\omega)$ versus
	$\abs{\omega}$. Here $E_\infty=\Tr(H_{\rm sec})/d_{\rm sec}$. In panel (b),
	$\abs{\bar E-E_\infty}\leq1.5J$,
	$D_{\rm mc}$ counts states satisfying $\abs{E-E_\infty}\leq1.5J$, and the
	corresponding counts are $D_{\rm mc}=10,37,126,461$. The frequency-bin width
	is $0.5J$, and the plotted range is $0\leq\abs{\omega}\leq6J$. Bins
	containing fewer than 12 unordered
	pairs are omitted. Error bars are standard errors of the empirical mean of
	$\abs{(O_c)_{nm}}^2$. The factor $D_{\rm mc}$ is a finite-window entropy
	proxy and should not be equated with $e^{S(\bar E)}$ at these sizes.}
	\label{fig:IsingcomponentCH}
\end{figure}

Figure~\ref{fig:IsingcomponentCH}(a) shows that the diagonal eigenstate
expectation values for $N=6,8,10,12$ organize around a common smooth dependence
on energy density, with visible finite-size scatter. The central cloud becomes
clearly tighter between $N=8$ and $N=12$, although the approach is not
monotonic at the two smallest sizes. Panel (b) retains the frequency dependence
of the off-diagonal envelope. The larger sizes follow a similar nonconstant
envelope, whereas $N=6$ is visibly noisier. In the fixed bin
$0.5\leq\abs{\omega}<1$, the values of $D_{\rm mc}V_{O_c}$ are $0.510$,
$0.571$, $0.497$, and $0.456$ for $N=6,8,10,12$, respectively, while
$D_{\rm mc}=10,37,126,461$. The corresponding unscaled root-mean-square
matrix elements decrease from $0.226$ to $0.0314$. This four-size near collapse
is consistent with $V_{O_c}\propto D_{\rm mc}^{-1}$ for this observable,
energy shell, and frequency bin, although visible size dependence remains at
the lowest frequencies. It does not establish a thermodynamic scaling law:
the calculation still concerns one fixed-support operator and one symmetry
sector at finite sizes
\cite{DAlessio:2015qtq,Beugeling:2015,Mondaini:2017}.

\FloatBarrier

To examine unitary dynamics, we instead use the intensive magnetization
$m_x=S_x/N$, where $S_x=\sum_{i=1}^N\sigma_i^x$, and compute
\be
\langle m_x(t)\rangle
=\bra{Y+}e^{iHt}m_xe^{-iHt}\ket{Y+},
\qquad
\ket{Y+}=\bigotimes_{i=1}^N\ket{Y+}_i,
\ee
where $\sigma_i^y\ket{Y+}_i=\ket{Y+}_i$.

\begin{figure}[!ht]
	\centering
\includegraphics[width=0.45\linewidth]{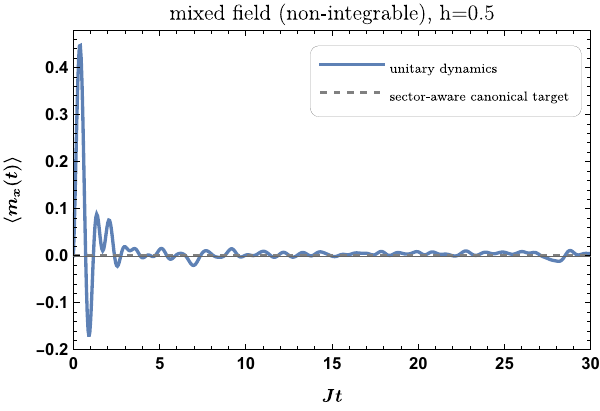}
\includegraphics[width=0.45\linewidth]{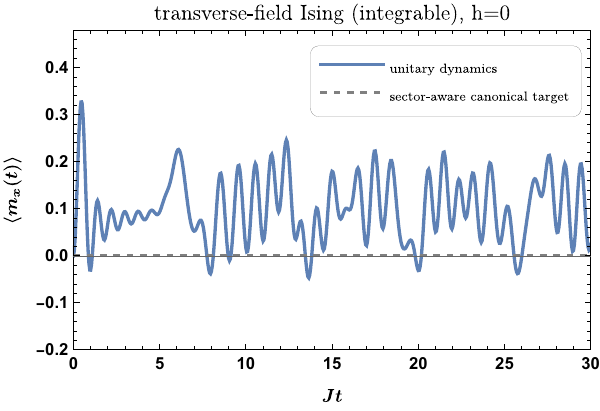}
	\caption{Exact symmetry-resolved unitary 
    evolution of $m_x=S_x/N$ from the
	homogeneous product state $\ket{Y+}$ in the open chain at $N=12$, $J=1$,
	and $g=-1.05$, over $0\leq Jt\leq30$ with time step
	$\Delta(Jt)=0.05$. Left: the non-integrable mixed-field chain at $h=0.5$
	in the reflection-even block ($R=+1$, $d_{\rm sec}=2080$). Right: the
	integrable transverse-field Ising chain at $h=0$ in the reflection-even,
	spin-flip sectors $F=+1$ and $F=-1$, of dimensions $1056$ and $1024$.
	Both panels use the same vertical scale. Dashed lines are energy-matched
	sector-aware canonical references. At $h=0$, the canonical density matrix is
	normalized separately in each occupied spin-flip sector at a common inverse
	temperature and recombined with the conserved initial sector weights, which
	are $1/2$ in each sector for this state. These
	finite-size curves illustrate rapid damping versus persistent structured
	oscillations; they do not establish a thermodynamic-limit relaxation rate.}
	\label{fig:EV-sigmaz}
\end{figure}

Figure~\ref{fig:EV-sigmaz} compares the two Hamiltonian branches on a common
vertical scale. In the mixed-field chain, the initial transient is rapidly
damped and $\langle m_x(t)\rangle$ subsequently exhibits only small
fluctuations around its energy-matched canonical reference. In the integrable
chain, substantially larger structured oscillations persist throughout the
displayed interval. Over $5\leq Jt\leq30$, the root-mean-square deviations
from the respective canonical references are approximately $0.0060$ and
$0.121$ in the mixed-field and integrable branches. The canonical line in the
integrable panel is only a sector-aware comparison reference; it is not a
generalized Gibbs ensemble containing the full set of integrable conserved
charges.

This example demonstrates rapid dephasing of one intensive observable in a
non-integrable finite chain under a single Hamiltonian, without averaging over
Hamiltonians. It does not imply that every observable or initial state
thermalizes. The initial energy distribution, effective dimension,
conserved-sector weights, recurrence scale, and overlaps with atypical
eigenstates remain relevant. Section~\ref{sec:weak} examines this dependence
directly using reduced density matrices.

\FloatBarrier

%%%%%%%%%%%%%%%%%%%%%%%%%%%%%%%%%%%%%%%%%%
%%%%%%%%%%%%%%%%%%%%%%%%%%%%%%%%%%%%%%%%%%
\section{The Role of Initial State: Weak and Strong Thermalization}
\label{sec:weak}

Non-integrability alone does not guarantee thermalization for every observable and initial state. The energy distribution, effective dimension, conserved-sector weights, and overlaps with atypical eigenstates also matter. Following Ba\~nuls, Cirac, and Hastings \cite{Banuls:2011vuw}, we call thermalization \emph{dynamically strong} when a fixed finite subsystem approaches its thermal reduced state directly, and \emph{dynamically weak} when the instantaneous reduced state can retain appreciable oscillations but an appropriate time average approaches the thermal state.

Let $\rho_A(t)=\Tr_B\rho(t)$, with $B$ the complement of $A$, and define $\rho_{A,\rm th}=\Tr_B\rho_{\rm th}$. The standard trace distance is
\be
D_A(t) = \frac{1}{2}\norm{\rho_A(t) - \rho_{A,\rm th}}_1,
\qquad \norm{X}_1 = \Tr\sqrt{X^\dagger X}.
\ee
For the weak notion, introduce the late-time running average used in Ref.~\cite{Banuls:2011vuw}:
\be
\overline{\rho}_A(t) = \frac{2}{t}\int_{t/2}^{t} d\tau\,\rho_A(\tau),
\qquad
\overline D_A(t) = \frac{1}{2}\norm{\overline{\rho}_A(t) - \rho_{A,\rm th}}_1.
\ee
Strong thermalization means that $D_A(t)$ becomes small after relaxation in the thermodynamic limit, whereas weak thermalization requires only $\overline D_A(t)\to 0$ while $D_A(t)$ may continue to oscillate. The order of limits matters: one may choose a sequence of observation times $t_N$ satisfying $t_{\rm rel}(N)\ll t_N\ll t_{\rm rec}(N)$ and then take $N\to\infty$ at fixed subsystem $A$. An infinite-time pointwise limit at fixed finite $N$ is not intended, because finite systems have recurrences. These dynamical notions must not be confused with strong and weak versions of ETH discussed in Section~\ref{sec:high}.

To investigate the dependence on the initial state, we restrict attention to
the homogeneous product-state family
\be\label{initial}
\ket{\theta,\phi}
=\bigotimes_{i=1}^{N}\left(\cos\frac{\theta}{2}\ket{Z+}_i
+e^{i\phi}\sin\frac{\theta}{2}\ket{Z-}_i\right).
\ee
Equation~\eqref{initial} defines only a two-parameter, translationally invariant product-state manifold. A general site-dependent pure product state has $2N$ real parameters, whereas a general normalized pure state of $N$ qubits has $2^{N+1}-2$ real parameters after removing the global phase. The examples below therefore illustrate initial-state dependence only within this restricted preparation family; they do not classify arbitrary product states, let alone entangled states.

\begin{figure}[!ht]
	\centering
\includegraphics[width=0.45\linewidth]{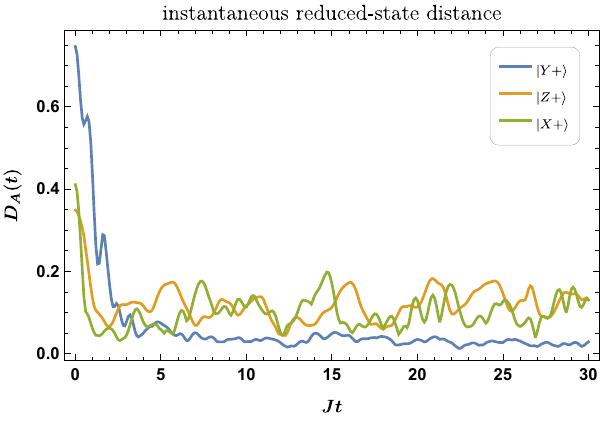}
\includegraphics[width=0.45\linewidth]{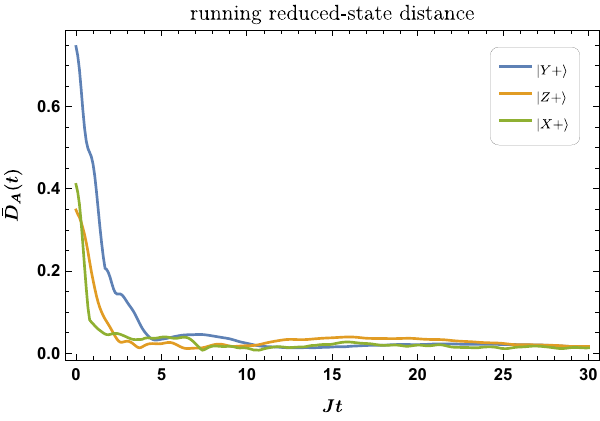}
	\caption{Initial-state dependence of 
    local relaxation in the non-integrable
	open Ising chain at $N=12$, $J=1$, $g=-1.05$, and $h=0.5$, in the
	reflection-even block ($R=+1$, $d_{\rm sec}=2080$). The subsystem is the
	central pair $A=\{6,7\}$, and the time interval is $0\leq Jt\leq30$.
	The exact unitary trajectories use $\Delta(Jt)=0.05$; reduced states and
	trace distances are sampled every $0.10$ in $Jt$. Left:
	$D_A(t)=\frac12\norm{\rho_A(t)-\rho_{A,\rm th}}_1$. Right:
	$\overline D_A(t)=\frac12\norm{\overline\rho_A(t)-\rho_{A,\rm th}}_1$,
	where $\overline\rho_A(t)=\frac{2}{t}\int_{t/2}^{t}\rho_A(\tau)d\tau$.
	Thus the right panel is the distance of the averaged reduced state, not the
	time average of $D_A(t)$. A separate energy-matched canonical reduced state
	is used for each of $\ket{Y+}$, $\ket{Z+}$, and $\ket{X+}$ within the
	reflection-even sector. The running integral is evaluated by a
	piecewise-linear trapezoidal approximation on the trace-distance grid.
	These are finite-$N$, finite-time diagnostics and do not
	establish asymptotic strong or weak thermalization.}
	\label{fig:1}
\end{figure}

The three preparations used in Figure~\ref{fig:1} are
$\ket{Z+}=\ket{0,0}$, $\ket{X+}=\ket{\pi/2,0}$, and
$\ket{Y+}=\ket{\pi/2,\pi/2}$. For each preparation, the inverse temperature
and canonical reduced state are determined separately by matching that
state's mean energy within the conserved reflection-even sector. The left
panel shows that the instantaneous two-site distance decreases strongly after
the initial transient but remains finite and fluctuating, especially for
$\ket{Z+}$ and $\ket{X+}$. The right panel is not a time average of
$D_A(t)$; it is the distance of the running-averaged reduced density matrix.
All three running distances become much smaller over the displayed window,
reaching approximately $0.013$--$0.017$ at $Jt=30$. Thus temporal averaging
strongly suppresses the finite-size coherent fluctuations for all three
states. The result is consistent with stronger direct relaxation for
$\ket{Y+}$ and a larger role for temporal averaging for $\ket{Z+}$ and
$\ket{X+}$, but it does not establish pointwise convergence or a
thermodynamic-limit classification.

\FloatBarrier

One useful covariate is the effective inverse temperature $\beta$, fixed by matching the mean energy:
\be\label{beta0}
\Tr(\rho_{\rm th}H) = E_0,
\ee
where $E_0 = \Tr(\rho_0 H)$. At finite size, if the initial state lies in one irreducible symmetry sector, the thermal trace must be restricted to that sector. If it occupies a direct sum of conserved sectors, one must instead form a separately normalized canonical state in every occupied sector and combine them with the fixed initial sector weights; a common $\beta$ is determined by the total mean energy. The code appendices follow this convention.

For the particular Ising model and homogeneous-product-state family studied in Ref.~\cite{Banuls:2011vuw}, states near $\beta=0$ were empirically associated with more strongly damped dynamics, while some low-temperature states exhibited weak thermalization. This is neither necessary nor sufficient in general. Writing $p_n = \abs{\braket{E_n}{\psi_0}}^2$, other relevant data include the energy variance $\sigma_E^2 = \sum_n p_n(E_n-E_0)^2$, the effective dimension $d_{\rm eff} = 1/\sum_n p_n^2$, conserved-sector weights, and overlaps with atypical eigenstates. The mean energy fixes the thermal target but not the rate of approach. For reference, define the normalized spectral position
\be\label{NE}
\mathcal{E} = \frac{\Tr(\rho_0 H) - E_{\min}}{E_{\max} - E_{\min}},
\ee
where $E_{\max}$ and $E_{\min}$ are the spectral extrema of the same resolved sector used for the state. In one sector $\alpha$, infinite temperature corresponds to $E_\infty^{(\alpha)}=\Tr_\alpha(H_\alpha)/\mathcal{D}_\alpha$; if the initial state occupies several conserved sectors with weights $w_\alpha$, the appropriate reference is $E_\infty=\sum_\alpha w_\alpha E_\infty^{(\alpha)}$. It is not universally located at $\mathcal{E}=1/2$, and tracelessness in the full Hilbert space does not by itself imply a zero trace in every symmetry block. Because the spin spectrum is bounded, negative temperatures are also possible, and $\abs{\beta}$ measures distance from infinite temperature rather than the sign of the energy. A quasiparticle interpretation can apply near a dilute low-energy regime, where a few coherent frequency scales dephase slowly \cite{Lin:2016egw}; it is model-dependent, and later decay is not excluded.

To explore this point further, consider the spin-$1/2$ Ising model:
\bea\label{Ising}
H &=& \sum_{i=1}^{N-1}\left(J_x\sigma^x_i\sigma^x_{i+1} + J_y\sigma^y_i\sigma^y_{i+1} + J_z\sigma^z_i\sigma^z_{i+1}\right) \nonumber\\
&&+ \sum_{i=1}^{N}\left(h_x\sigma^x_i + h_y\sigma^y_i + h_z\sigma^z_i\right)
+ g_x\sigma^x_1 + g_y\sigma^y_1 + g_z\sigma^z_1.
\eea
The couplings $(J_i, h_i, g_i)$ specify the model; chaos or integrability must be diagnosed after resolving its symmetries and cannot be read from nonzero couplings alone. With $J_z=-J$, $h_x=-g$, $h_z=-h$, and all other couplings zero, Eq.~\eqref{Ising} reduces to Eq.~\eqref{Ising0}. For the parameters studied in Ref.~\cite{Banuls:2011vuw}, the homogeneous states $\ket{X+}$, $\ket{Y+}$, and $\ket{Z+}$ show visibly different finite-time dynamics: $\ket{Y+}$ is strongly damped, while $\ket{Z+}$ and $\ket{X+}$ show weaker damping. The long-lived $\ket{X+}$ oscillations have substantial finite-size effects and are not evidence of permanent nonthermalization \cite{Sun:2020ybj, Banuls:2011vuw}.

For the general initial state \eqref{initial} and the model \eqref{Ising}, the expectation value of the energy takes the form
\be\label{EE}
\begin{aligned}
E ={}& (N-1)\Bigl[\sin^2\theta\bigl(J_x\cos^2\phi + J_y\sin^2\phi\bigr) + J_z\cos^2\theta\Bigr] \\
&+ N\Bigl[\sin\theta\bigl(h_x\cos\phi + h_y\sin\phi\bigr) + h_z\cos\theta\Bigr] \\
&+ \sin\theta\bigl(g_x\cos\phi + g_y\sin\phi\bigr) + g_z\cos\theta.
\end{aligned}
\ee

\begin{figure}[!ht]
	\begin{center}
		\includegraphics[width=0.55\linewidth]{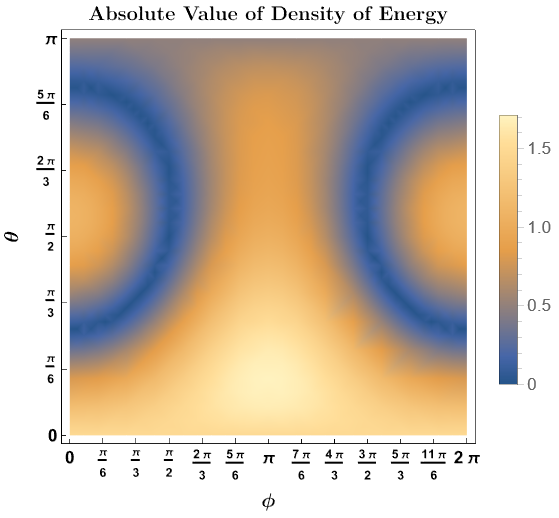}
	\end{center}
	\caption{Absolute energy density from Eq.~\eqref{EE} at $N=100$ for the restricted homogeneous-product-state family. For this traceless Hamiltonian, dark regions lie near the infinite-temperature energy. They are candidate regions for strongly damped dynamics, not a stand-alone thermalization diagnostic.}
	\label{fig:E-Ising1}
\end{figure}

Figure~\ref{fig:E-Ising1} plots $\abs{E}/N$ from Eq.~\eqref{EE} at $N=100$ for $J_z=-1$, $h_x=1.05$, and $h_z=-0.5$, with all other couplings zero. This mixed-field model is non-integrable after symmetry resolution. Because the Hamiltonian is traceless, dark regions lie close to the infinite-temperature energy and are therefore \emph{candidates} for more strongly damped dynamics within the two-parameter family. Energy density alone does not establish strong thermalization or identify a unique relaxation rate; states with the same mean energy can have different variances, effective dimensions, and atypical-state overlaps. Other diagnostics are discussed in Refs.~\cite{Sun:2020ybj, Chen:2021, Lin:2016egw, Prazeres:2023hce, Bhattacharjee:2022qjw, Nandy:2023brt, Alishahiha:2024rwm}.

\begin{figure}[!ht]
	\begin{center}
		\includegraphics[width=0.55\linewidth]{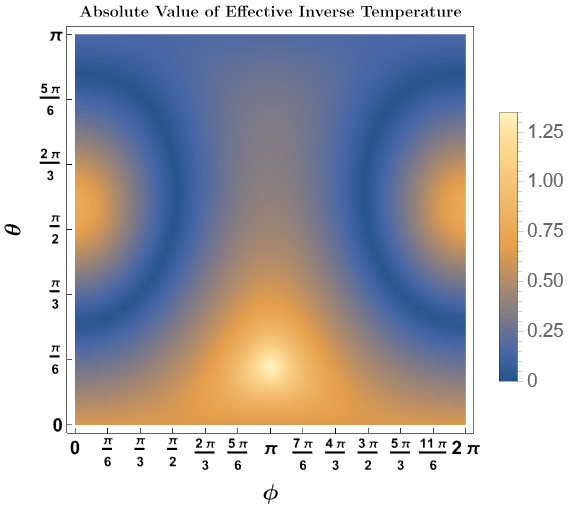}
	\end{center}
	\caption{Absolute effective inverse temperature for the homogeneous product states in Eq.~\eqref{initial}, at $N=9$ with $J_z=-1$, $h_x=1.05$, and $h_z=-0.5$. The signed $\beta$ is determined by Eq.~\eqref{beta0}; the plot displays its magnitude.}
	\label{fig:beta}
\end{figure}

Figure~\ref{fig:beta} shows $\abs{\beta}$ obtained by solving Eq.~\eqref{beta0} at $N=9$. Its qualitative correlation with $\abs{E}/N$ is expected because the finite-size canonical energy is a monotone function of $\beta$, but the nonlinear map and finite-size effects prevent a direct quantitative comparison between the $N=9$ and $N=100$ panels. No scaling conclusion or percentage error follows from these two plots alone.

\FloatBarrier

Figures~\ref{fig:E-Ising1} and~\ref{fig:beta} should therefore be read as maps of thermodynamic covariates, not as classifications by themselves. The labels in Figure~\ref{fig:1} come from the actual time-dependent observable (or reduced-state) behavior. Any apparent association with $E$ or $\beta$ is empirical for this model, system-size range, and homogeneous-product-state manifold. These figures cannot support conclusions about arbitrary site-dependent product states or entangled initial states.

%%%%%%%%%%%%%%%%%%%%%%%%%%%%%%%%%%%%%%%%%
%%%%%%%%%%%%%%%%%%%%%%%%%%%%%%%%%%%%%%%%%

\section{Limits of ETH and High-Energy Connections}
\label{sec:high}

\subsection{Strong and Weak ETH; Mechanisms of Violation}

The dynamical terminology of Section~\ref{sec:weak} 
is distinct from strong and weak ETH. Strong ETH states 
that, within a specified bulk energy-density window 
and irreducible symmetry sector, every eigenstate 
becomes locally thermal as $N\to\infty$. Weak ETH 
requires only that the fraction of nonthermal 
eigenstates in that window vanishes in the 
thermodynamic limit—that is, the measure of
exceptional states goes to zero as $N\to\infty$,
even though a sparse set of atypical eigenstates may 
persist. Spectral edges are normally outside the
window, so their exceptional behavior does not by 
itself disprove strong ETH as just defined. 
Neither version requires a non-degenerate spectrum; 
exact symmetries and protected degeneracies must 
instead be resolved. The non-degenerate-gap assumption 
entered only the elementary fluctuation formula in 
Section~\ref{sec:equ}.

Thermalization requires more than matching mean energy. For the selected local or few-body class, one needs $\Tr[(\rho_{\rm DE}-\rho_{\rm MC})O]\to 0$, or convergence of finite-subsystem reduced states. OTOCs diagnose scrambling and operator growth, while level statistics diagnose spectral correlations; neither alone demonstrates thermalization. For quench dynamics, one must additionally specify the initial energy distribution, conserved-sector weights, and effective dimension.

Conventional ETH generally fails in integrable and many-body-localized phases, but the reason is not an ordinary finite symmetry. Integrable models possess extensive local or quasilocal constraints and are described, where appropriate, by generalized Gibbs ensembles (GGEs); many-body-localized systems possess an extensive set of emergent local integrals of motion \cite{Basko:2006, Serbyn:2013, Huse:2014}. Thermalization can still be analyzed sector by sector in the presence of ordinary symmetries. Systematic counterexamples to ETH and quenches that thermalize without satisfying its strongest form are discussed in Refs.~\cite{Shiraishi:2017, Mori:2017}.

Quantum many-body scars offer a distinct mechanism for ETH violation. The constrained PXP model provides a paradigmatic example: after resolving all spatial symmetries, the model is non-integrable and its spectrum is largely thermal, yet it contains a sparse set of atypical, low-entanglement eigenstates. These special states produce long-lived periodic revivals for certain initial states, a phenomenon observed experimentally in Rydberg atom chains \cite{Bernien:2017, Lesanovsky:2012, Turner:2018P, Turner:2018N}. The term "scar" is borrowed from single-particle quantum chaos \cite{Heller:1984}, where it refers to eigenstates concentrated near unstable classical periodic orbits. In the many-body setting, however, this classical orbit analogy is only one possible organizing principle; it should not be taken as a universal definition. Many-body scars violate strong ETH in the bulk of the spectrum, but they remain compatible with weak ETH provided their fraction vanishes in the thermodynamic limit. These states have now been identified in a variety of quantum many-body models and have found practical applications in quantum information, including the preparation and protection of entangled states \cite{Omran:2019unq, Serbyn:2020wys}. For a deeper exploration of the origins and classification of Hilbert-space scars, see Ref.~\cite{Guo:2023wck}.

It is important to distinguish quantum many-body scars from a separate mechanism of ETH breaking: Hilbert-space fragmentation \cite{Khemani:2019vor, Sala2020}. In fragmented systems, local constraints divide the Hilbert space into dynamically disconnected Krylov sectors, preventing the system from exploring the full Hilbert space under unitary evolution. A particularly strong form, often called "Hilbert-space shattering," occurs when the number of such sectors grows exponentially with system size. However, this term should not be used as an automatic synonym for every fragmented model; the size and number of fragments can vary, and individual sectors need not all be small. In fragmented systems, ETH fails across the full Hilbert space, although sufficiently large individual fragments may still thermalize after sector resolution. Scars, by contrast, do not require fragmentation: they are atypical eigenstates embedded within an otherwise thermalizing sector. This distinction is crucial for interpreting experimental revivals and for applications in quantum information, where scars offer a route to preparing and protecting nonthermal states \cite{Omran:2019unq, Serbyn:2020wys}.

%%%%%%%%%%%%%%%%%%%%%%%%%%%%%%%%%%%%%

\subsection{Connections to High-Energy Physics}

ETH has important applications in holographic duality and black-hole physics. In the AdS/CFT correspondence \cite{Maldacena:1997re, Aharony:1999ti}, a thermal state in the boundary field theory corresponds to a black hole in the bulk gravitational theory. At high temperatures, the dominant bulk configuration is a black hole; below a critical temperature—the Hawking--Page transition—thermal AdS space takes over. A heavy pure eigenstate in the field theory can be described by a semiclassical bulk geometry, but only through coarse-grained boundary observables; the fine-grained quantum state remains pure. Holographic quenches provide a powerful framework for studying thermalization: the process of black-hole formation and subsequent relaxation in the bulk directly models thermalization and hydrodynamization of the boundary theory. Black-hole evaporation, by contrast, is a separate phenomenon. It concerns the recovery of fine-grained quantum information and should not be conflated with thermalization itself.

In large-$N$, strongly coupled holographic CFTs, the growth of OTOCs and the butterfly effect can be described geometrically by shock waves near the black hole horizon \cite{Shenker:2013pqa, Shenker:2013yza, Leichenauer:2014nxa, Roberts:2014isa}. The idea is simple: a small perturbation near the horizon grows over time due to gravitational backreaction, eventually forming a shock wave that propagates along the horizon. In the dual field theory, this shock wave corresponds to the butterfly effect—the rapid spread of quantum information. This geometric description relies on the holographic and semiclassical limits, so it does not apply to every quantum field theory. Nevertheless, it provides a powerful connection between gravitational backreaction, operator growth, and information scrambling \cite{Maldacena:2015waa}.

The Sachdev--Ye--Kitaev (SYK) model offers a concrete playground for testing these ideas \cite{Kitaev:2015, Sachdev:2015efa, Maldacena:2016hyu}. At large $N$, strong coupling, and low temperature, the model's low-energy dynamics is described by the Schwarzian action. In this regime, the Lyapunov exponent saturates the universal bound $\lambda_L \leq 2\pi/\beta$, known as the MSS bound. Remarkably, the same Schwarzian dynamics also appears in Jackiw--Teitelboim (JT) gravity, a two-dimensional theory of gravity \cite{Jensen:2016pah, Maldacena:2016upp, Engelsoy:2016xyb, Almheiri:2014cka}. The SYK model has been shown to satisfy ETH for typical disorder realizations \cite{Sonner:2017hxc}, but one should be cautious: not every finite-$N$ or SYK-like model exhibits maximal chaos or satisfies ETH.

An even more surprising connection emerges at the nonperturbative level. JT gravity is not dual to a single quantum system but rather to an ensemble of random matrices \cite{Saad:2019lba}. In this formulation, the gravitational path integral computes averages over this ensemble. Thus JT gravity provides a tractable model of universal spectral correlations, but it is not a universal model of quantum chaos applicable to all systems.

As discussed in Section~\ref{sec:ETH}, the MSS bound $\lambda_L \leq 2\pi/\beta$ holds under general assumptions of thermal analyticity and factorization \cite{Maldacena:2015waa}. Semiclassical black holes and strongly coupled large-$N$ SYK saturate this bound, but generic systems need not. The holographic computation of OTOCs has revealed a deep link between black hole geometry, quantum chaos, and thermalization. Holographic quenches further illustrate this connection by modeling how far-from-equilibrium states relax and hydrodynamize \cite{Balasubramanian:2011ur, Das:2010yw, Bellantuono:2016mhl, Chen:2023iws}. However, it is important to remember that chaos, scrambling, and thermalization are distinct concepts; they are related but not synonymous.

%%%%%%%%%%%%%%%%%%%%%%%%%%%%%%%%%%%%%%%%%%%%%%
\section{Conclusions and Outlook}
\label{sec:outlook}

Equilibration and thermalization are distinct phenomena. Dephasing can make local observables stay close to their diagonal-ensemble values for most late times. But thermalization goes further: it requires that these values match those of a thermal ensemble, such as the microcanonical or canonical ensemble, for the chosen class of local or few-body observables.

In a finite isolated system, the expectation value $\langle O(t)\rangle$ is generally quasiperiodic. It need not converge to a constant at every instant. Both pure and mixed initial states can be described using the same density-matrix framework. However, their energy populations and coherences affect the fluctuation amplitudes and the relaxation behavior. Equilibration to the diagonal ensemble can occur when the relevant off-diagonal matrix elements are small, when dephasing is effective, and when the initial state has sufficiently broad support in energy. None of these conditions, by themselves, make the global state thermal.

Thermalization is a special case of equilibration. It occurs when the diagonal ensemble and the thermal ensemble become indistinguishable for the observables of interest. Concretely, this means $\Tr[(\rho_{\rm DE}-\rho_{\rm MC})O]\to 0$ for every selected local or few-body observable $O$, or equivalently, that the reduced states of finite subsystems converge to their thermal forms. Matching the mean energy is necessary, but it is not sufficient for thermalization.

ETH provides the required structure. It ensures that the diagonal matrix elements $O_{nn}$ vary smoothly with energy within a bulk energy window, and that the off-diagonal elements are exponentially suppressed by the entropy factor. Strong ETH states that every eigenstate in such a window has thermal local properties in the thermodynamic limit. This does not mean that a global pure eigenstate is identical to a mixed thermal density matrix; the two are fundamentally different objects.

ETH itself does not require a non-degenerate spectrum. Exact symmetries and protected degeneracies must be resolved, and the analysis should be carried out within each symmetry sector. The stronger assumption of non-degenerate energy gaps is used only for the simple fluctuation identity in Section~\ref{sec:equ}. To conclude thermalization after a quench, the initial-state sequence must have a narrow energy-density distribution and appropriate weights in each conserved sector. The observable must also belong to the stated local or few-body class.

Various diagnostics can probe different aspects of thermalization. Entanglement measures test whether local subsystems have thermal properties. OTOCs diagnose scrambling and operator growth. While these phenomena are related in certain models, no single diagnostic—whether an OTOC or a level-spacing statistic—is by itself sufficient to demonstrate thermalization.

The Ising-chain examples presented in this primer illustrate, at finite size, how symmetry-resolved spectra, operator matrix elements, and restricted families of initial states probe these ideas. Dynamically strong and weak thermalization depend on more than mean energy: the energy variance, effective dimension, sector weights, transport properties, and overlaps with atypical eigenstates all play important roles \cite{Lin:2016egw, Prazeres:2023hce, Alishahiha:2024rwm}. These examples also highlight the importance of careful numerical analysis, including the resolution of all symmetries, the use of appropriate thermal references, and the need for controlled finite-size scaling.

Several important extensions and open directions deserve mention.

\begin{enumerate}

\item \textbf{Conditions for ETH.}
The precise conditions under which ETH holds in generic non-integrable systems remain to be fully understood. While ETH is well supported by numerical evidence in many models, a complete characterization of when it applies—and when it fails—is still lacking. The development of rigorous mathematical results for ETH, beyond the heuristic and numerical evidence, is an important goal. While ETH is widely accepted in the physics community, a full mathematical proof or a precise set of necessary and sufficient conditions would significantly deepen our understanding of quantum thermalization.

\item \textbf{The role of initial states.}
The role of initial states in thermalization remains a rich area of investigation. The distinction between dynamically strong and weak thermalization, and the dependence on energy distribution and effective dimension, call for a more complete classification of initial states and their thermalization properties. Not all initial states thermalize at the same rate, and some may fail to thermalize altogether due to special structures such as scars or fragmentation. Recent progress using the Krylov basis and complexity measures offers new perspectives on how initial states relax and how thermalization proceeds in the presence of conserved quantities \cite{Alishahiha:2024rwm}. These approaches may eventually provide a unified framework for understanding the dynamics of thermalization from a information-theoretic perspective.

\item \textbf{Symmetries and sector-by-sector analysis.} 
Ordinary symmetries must be treated with care. In systems with such symmetries, the analysis must be carried out sector by sector, and ETH should be formulated within each irreducible representation \cite{Fukushima:2025, Murthy:2023}. More exotic symmetries—such as higher-form or non-Abelian symmetries—introduce additional structure that can modify thermalization and may require generalizations of the standard framework. Understanding how different types of symmetries affect ETH remains an active and important direction.

\item \textbf{Diagnostics of quantum chaos.}
The relationship between spectral correlations, eigenstate properties, and dynamical observables is still being clarified. Recent work suggests that these different diagnostics probe different aspects of quantum chaos. A deeper understanding of how these diagnostics are connected, and when they agree or disagree, is an important goal for the future.

\item \textbf{Disordered systems and many-body localization.}
Disordered systems and many-body localization (MBL) represent a dramatic breakdown of ETH. In the MBL phase, an extensive set of emergent local integrals of motion prevents thermalization, and eigenstates violate ETH in a strong sense \cite{Sierant:2025, Nandkishore:2015, Abanin:2019}. The transition between the thermal and MBL phases remains an active research area. Key open questions include the nature of the transition itself, the role of rare regions, and whether the MBL phase is stable in the thermodynamic limit. These questions touch upon fundamental issues in quantum statistical mechanics and continue to attract significant theoretical and experimental attention.

\item \textbf{Open quantum systems.}
Open quantum systems require a different approach from the closed-system ETH ansatz. Unlike closed systems, open systems exchange energy and information with an environment. Stationary behavior arises through dissipation and decoherence, and must be characterized using the generator of the open dynamics and the resulting steady state. Recent work has explored the universality, robustness, and limits of ETH-like statements in open systems \cite{Almeida:2026, Shirai:2020}. These studies reveal new phenomena, such as entanglement phase transitions and dynamical transitions in the spectral properties of the Liouvillian. However, these considerations do not by themselves imply such transitions in every open system; careful analysis is required on a case-by-case basis.

\item \textbf{Holographic models and quantum gravity.}
Holographic models offer a powerful setting for studying ETH and quantum chaos. The AdS/CFT correspondence relates thermalization in strongly coupled field theories to black-hole formation and relaxation in the bulk \cite{Jahnke:2018off}. The SYK model and JT gravity provide tractable examples where maximal chaos and ETH-like behavior can be studied in the appropriate limits: large $N$, strong coupling, low temperature, and semiclassical gravity. These models also reveal deep connections between random-matrix theory, quantum gravity, and the statistical mechanics of black holes. Future work may uncover further links between quantum information, geometry, and thermalization, potentially shedding light on the black-hole information paradox and the nature of spacetime itself.

\item \textbf{Thermalization and quantum information.}
The interplay between thermalization and quantum information continues to motivate new research directions. Examples include quantum simulation, where ETH helps predict the behavior of analog quantum simulators; quantum error correction, where thermalization can affect the stability of encoded information; and the black-hole information paradox, where ETH and related ideas may shed light on how information is preserved in quantum gravity. These connections promise to enrich both fields and may lead to new insights into the nature of quantum many-body dynamics.

\end{enumerate}

Throughout these diverse settings, several general lessons emerge.
First, claims of thermalization must always be precise. One must specify the observable class, the symmetry sector, the initial-state sequence, and the thermodynamic limit. Equilibration to the diagonal ensemble is a necessary step, but it is not sufficient for thermalization. The ETH provides a powerful sufficient framework, but its applicability must be checked on a case-by-case basis. Deviations from ETH—whether due to integrability, localization, scars, or fragmentation—are not merely failures of the framework; they reveal important physics about the structure of quantum many-body systems.

In summary, the Eigenstate Thermalization Hypothesis provides a remarkably successful framework for understanding thermalization in isolated quantum systems. It explains why individual eigenstates can appear thermal, why suitable initial states relax to thermal equilibrium, and why the diagonal ensemble captures the late-time behavior of local observables. While ETH is not universal—it fails in integrable, localized, and certain fragmented systems—its domain of validity covers a wide range of physically relevant models. The study of ETH and quantum thermalization remains a vibrant and rapidly evolving field. The concepts and tools presented in this primer—equilibration, thermalization, ETH, quantum chaos diagnostics, and the role of initial states—form the foundation for understanding how isolated quantum systems reach thermal equilibrium. As quantum simulators and processors continue to grow in size and capability, these ideas will become increasingly important for interpreting and predicting the behavior of complex quantum many-body systems.

%%%%%%%%%%%%%%%%%%%%%%%%%%%%%%%%%%%%%%%
%%%%%%%%%%%%%%%%%%%%%%%%%%%%%%%%%%%%%%%
\section*{Acknowledgments}

We would like to thank Komeil Babaei Velni, Souvik Banerjee, Ali Mollabashi, Mohammad Reza Mohammadi Mozaffar, Reza Pirmoradian, Mohammad Reza Tanhayi, Nilofar Vardian, and Hamed Zolfi for discussions on various aspects of quantum chaos and thermalization, and for their comments on the draft. This work was supported by the Iran National Science Foundation (INSF) under project No. 4023620. The authors also acknowledge the use of DeepSeek as an editorial assistant in improving the clarity and presentation of the manuscript.

%%%%%%%%%%%%%%%%%%%%%%%%%%%%%%%%%%%%%%%%%
%%%%%%%%%%%%%%%%%%%%%%%%%%%%%%%%%%%%%%%%

\appendix 

\section{Normalization of Level Spacing and Gap Ratio Statistics}
\label{app:level}

\subsection{Level spacing}

Level-spacing distributions are useful spectral diagnostics of integrability and quantum chaos, but are not sufficient by themselves to establish either ETH or thermalization. This appendix defines the normalization used below and introduces the adjacent-gap ratio, which does not require unfolding.

Before any level statistic is computed, the Hamiltonian must be block-diagonalized with respect to all unitary symmetries, and anti-unitary symmetries and Kramers degeneracies must be treated consistently. Eigenvalues from independent blocks must never be mixed. If a resolved block contains too few levels, level statistics are inconclusive; this does not make the model trivially integrable.

Consider one resolved block containing ${\cal D}_0\leq{\cal D}$ ordered energies $E_1<\cdots<E_{{\cal D}_0}$. It has ${\cal D}_0-1$ spacings $s_\alpha=E_{\alpha+1}-E_\alpha$. Their scale varies with the density of states, so an unfolding prescription is required before comparison with a universal spacing distribution.

For an interval containing ${\cal N}$ ordered levels, $\Delta E=E_{\cal N}-E_1$, there are ${\cal N}-1$ spacings. The corresponding mean density and spacing are
\be
d(E)=\frac{{\cal N}-1}{\Delta E}\,,
\qquad
s_{\rm av}=\frac{1}{{\cal N}-1}\sum_{\alpha=1}^{{\cal N}-1}s_\alpha
=\frac{E_{\cal N}-E_1}{{\cal N}-1}=\frac{1}{d(E)}\,.
\ee
To remove this local scale one defines
\be
\hat{s}_\alpha=d(E) s_\alpha,
\ee
so that the average of level spacing is one. Then, the distribution should be computed for $S_\alpha=\frac{\hat{s}_\alpha}{{\hat s}_{\rm av}}$, which is believed to exhibit universal features.

The preceding illustration assumed a constant density within the interval. In general the local density depends on the central energy, $d(E_\alpha)$, so an explicit local unfolding prescription must be stated.

Although unfolding is not unique, a convenient prescription \cite{Evnin:2018jbh} uses the interval from $E_{\alpha-\Delta}$ to $E_{\alpha+\Delta}$. It contains $2\Delta$ gaps (and $2\Delta+1$ levels), giving
\be
d(E_\alpha)=\frac{2\Delta}{E_{\alpha+\Delta}-E_{\alpha-\Delta}}\,.
\ee
One should also provide a prescription for determining $\Delta$. Generally, it could be any number proportional to a fractional power of the total number of energy eigenvalues ${\cal D}_0$. Usually, it is set to be the largest integer number smaller than $\sqrt{{\cal D}_0}$. Moreover, for this definition to make sense, one needs to assume $\alpha-\Delta\geq 1$ and $\alpha+\Delta\leq {\cal D}_0$, so that $\alpha=\Delta+1,\cdots, {\cal D}_0-\Delta$. Interestingly, this procedure automatically removes the edge effects by removing modes near the edges of the spectrum. By making use of this local mean density of states, the unfolded level spacing and the average of level spacing are given by:
\be
\hat{s}_\alpha=d(E_\alpha) (E_{\alpha+1}-E_\alpha)=\frac{2\Delta (E_{\alpha+1}-E_\alpha)}{E_{\alpha+\Delta}-E_{\alpha-\Delta}},\;\;\;\;\;\;\;
\hat{s}_{\rm av}=\frac{2\Delta}{{\cal D}_0-2\Delta}\;
\sum_{\alpha=\Delta+1}^{{\cal D}_0-\Delta} \,\frac{ E_{\alpha+1}-E_\alpha}{E_{\alpha+\Delta}-E_{\alpha-\Delta}}\,.
\ee
The normalized level spacing is $S_\alpha=\hat{s}_\alpha/\hat s_{\rm av}$. Under the usual random-matrix assumptions, unfolded spectra in chaotic regimes are commonly compared with the appropriate Wigner--Dyson distribution, while many integrable regimes are compared with Poisson statistics. Exceptions exist, and neither behavior is a definition of chaos or integrability. The factor $2\Delta$ cancels from the normalized spacing.

\subsection{Gap Ratio Statistics}

As discussed above, the level spacing distribution requires spectral unfolding to remove the dependence on the local density of states. However, there is no unique unfolding scheme that applies equally well to all models; the procedure must be chosen carefully depending on the specific system under investigation. To circumvent this difficulty, Oganesyan and Huse~\cite{Oganesyan:2007} proposed the gap ratio statistics, which completely eliminates the need for unfolding since ratios of consecutive spacings are independent of the local density of states.

For each energy eigenvalue $E_n$, one defines the adjacent gap ratios
\be
\tilde{r}_n = \frac{\min(s_n,\, s_{n-1})}{\max(s_n,\, s_{n-1})}
= \min \left( r_n, \frac{1}{r_n} \right),
\ee
where $s_n = E_{n+1} - E_n$ are the consecutive level spacings and $r_n = s_n/s_{n-1}$.

Atas, Bogomolny, Giraud, and Roux~\cite{Atas:2013} derived Wigner-like surmises for all three classical random-matrix ensembles. For the folded ratio $0\leq\tilde r\leq1$ the probability density is
\be\label{ratioch}
P_W(\tilde{r}) = \frac{1}{Z_{\delta}}
\frac{(\tilde{r}+\tilde{r}^2)^{\delta}}
{(1+\tilde{r}+\tilde{r}^2)^{1+\tfrac{3}{2}\delta}},
\ee
where $\delta=1,2,4$ for GOE, GUE, and GSE. This folded density is twice the conventional Atas density for the unfurled ratio $r\in[0,\infty)$ evaluated at $r=\tilde r$. Accordingly, the constants quoted below are the folded-domain normalizations.

The average value of $\tilde{r}$ serves as a useful diagnostic for distinguishing between different spectral statistics. For an uncorrelated Poisson spectrum, the probability distribution takes the simple form
\be\label{ratioin}
P_{\text{Poisson}}(\tilde{r}) = \frac{2}{(1+\tilde{r})^2},
\ee
yielding $\langle \tilde{r} \rangle = 2\ln 2 - 1 \approx 0.386$. For the three classical random matrix ensembles, the averages are
\be
\langle \tilde{r} \rangle_{\text{GOE}} = 4-2\sqrt{3} \approx 0.536,
\ee
\be
\langle \tilde{r} \rangle_{\text{GUE}} = \frac{2\sqrt{3}}{\pi}-\frac{1}{2} \approx 0.602,
\ee
\be
\langle \tilde{r} \rangle_{\text{GSE}} = \frac{32}{15}\frac{\sqrt{3}}{\pi}-\frac{1}{2} \approx 0.676.
\ee
The corresponding normalization constants in Eq.~\eqref{ratioch} are $Z_1 = 4/27$ for GOE, $Z_2 = 2\pi/(81\sqrt{3})$ for GUE, and $Z_4 = 2\pi/(729\sqrt{3})$ for GSE. These values are widely used as convenient benchmarks to identify the spectral statistics of a given model.

The gap ratio has become the standard diagnostic for quantum chaos in many-body physics due to its simplicity and robustness, particularly for systems with small Hilbert space dimensions or complicated density of states where unfolding is unreliable \cite{Atas:2013}.

 \section{Reproducible Mathematica  implementation}
 \label{app:mathematica}
	
In this supplementary file, we present  Mathematica scripts utilized to obtain the numerical results presented in this paper. 
These scripts may serve as basic illustrations for readers who wish to acquaint themselves with performing calculations in Mathematica. One could also extend these Mathematica scripts for other models or 
to evaluate other quantities such as inverse participation
ratio.
	
	\subsection*{Hamiltonian and Initial state}
To proceed, we should first introduce the Hamiltonian and the initial state used in our numerical computations.
To do so, several preliminary definitions are required. First we define  $2\times 2$ identity matrix
		\begin{tcolorbox}[
		boxrule=0pt,
		sharp corners
		]
			\scriptsize
		\begin{lstlisting}[mathescape=true]
 identity = IdentityMatrix[2];
 
 sigmaAtN[mat_,n_,L_] := 
     Module[{list}, list = Table[identity, {L}];
     list[[n]] = mat;
     Fold[KroneckerProduct, First[list], Rest[list]]];
		\end{lstlisting}
	\end{tcolorbox}
 The second line in the above box shows how to set a  Pauli matrix at $n$-th position of a spin chain 
 with length $L$ ( $L$ sites).  By  making use of  this definition, one can introduce $ \sigma^x_n, \sigma^y_n $ and $\sigma^z_n$  as follows
	\begin{tcolorbox}[
	boxrule=0pt,
	sharp corners
	]
		\scriptsize
	\begin{lstlisting}[mathescape=true]
sigmax[n_,L_] := sigmaAtN[PauliMatrix[1], n, L];

sigmay[n_,L_] := sigmaAtN[PauliMatrix[2], n, L];

sigmaz[n_,L_] := sigmaAtN[PauliMatrix[3], n, L]; 
	\end{lstlisting}
\end{tcolorbox}
These are all we need to define a Hamiltonian 
for a generic Ising model involving $\sigma^{x,y,z}$. In particular, for the model we have 
considered in this paper 
\begin{eqnarray}
	H=-J \sum_{i=1}^{N-1} \sigma_i^z \sigma_{i+1}^z -\sum_{i=1}^{N}(g \sigma_{i}^x+ h  \sigma_{i}^z)
\end{eqnarray}
one write 
\begin{tcolorbox}[
	boxrule=0pt,
	sharp corners
	]
		\scriptsize
	\begin{lstlisting}[mathescape=true]
g = -105/100;
h = 1/2;
L=10;
 H= -g Sum[sigmax[i, L], {i, 1, L}] -     h Sum[sigmaz[i, L], {i, 1, L}] - 
      Sum[sigmaz[i, L].sigmaz[i + 1, L], {i, 1, L - 1}];
	\end{lstlisting}
\end{tcolorbox}
Additionally, it is imperative to define the initial state.  The initial state is a tensor product of individual single-qubit states at each site. Consequently, we start by defining the single qubit-state and then the tensor product state in the
following form 
\begin{tcolorbox}[
	boxrule=0pt,
	sharp corners
	]
	\scriptsize
	\begin{lstlisting}[mathescape=true]
singleState[theta_, phi_] :=
    Cos[theta/2]{{1}, {0}}+E^(I phi)Sin[theta/2] {{0}, {1}};
    
tensorProductState[stateVector_, n_] := 
    Fold[KroneckerProduct, stateVector, Table[stateVector, {n - 1}]];
	\end{lstlisting}
\end{tcolorbox}
By making use of these two definitions, one can introduce the general initial state as follows
\begin{tcolorbox}[
	boxrule=0pt,
	sharp corners
	]
	\scriptsize
	\begin{lstlisting}[mathescape=true]
initialState[theta_, phi_, L_] := 
 Flatten[tensorProductState[singleState[theta, phi], L]]
	\end{lstlisting}
\end{tcolorbox}

\subsection*{Level Spacing of Ising Model}
In Appendix A, we have demonstrated the method for computing the level spacing for generic models. Here we present the details of the computation of level spacing for the Hamiltonian \eqref{Ising0}. As previously stated, identifying the model's symmetries is crucial.
The Hamiltonian \eqref{Ising0} exhibits parity symmetry (see {\it e.g.} \cite{Joel:2013}). Within the Ising model framework, parity symmetry signifies that the system's properties are conserved under spatial inversion. For a one-dimensional Ising model, the parity operation, denoted by $\hat{\Pi}$, inverts the qubit sequence, mapping site $i$ to $ L + 1 - i $. For this model, reversing the spin order does not alter the system's energy. In other words, reversing a spin sequence does not affect the nearest-neighbor interactions, thereby maintaining the system's symmetry with respect to this parity operation.

To find out the parity of a state, we construct an orthogonal basis for the Hilbert space utilizing the quantum states of spin. These states are typically expressed using basis vectors that correspond to the eigenstates of $\frac{1}{2} \sigma_z$, where the spins oriented upwards in the z-direction are labeled as 
$|\uparrow \rangle =\begin{pmatrix}
	1	\\
	0
\end{pmatrix}$
, and those pointing downward are labeled as
$|\downarrow \rangle =\begin{pmatrix}
	0	\\
	1
\end{pmatrix}$
. Using this basis, we can easily investigate the parity of states.
. For example the parity pair of the state 
$|\downarrow \uparrow \downarrow \downarrow\downarrow \rangle$ 
is
$|\downarrow  \downarrow \downarrow \uparrow \downarrow \rangle$ . In other words, these two states are related to each other by the parity operator, 
$\hat{\Pi}|\downarrow \uparrow \downarrow \downarrow\downarrow \rangle=|\downarrow  \downarrow \downarrow \uparrow \downarrow \rangle$
.
According to these explanations, the first step is to build the  quantum states of the spins

\begin{tcolorbox}[
	boxrule=0pt,
	sharp corners
	]
	\scriptsize
	\begin{lstlisting}
   Upstate = Table[1, {i, 1, L}];
   Downstate = Table[0, {i, 1, L}];
  SPINQBasis = Permutations[Join[Upstate, Downstate], {L}];
	\end{lstlisting}
\end{tcolorbox}
In the next step, we find which two bases related together with the parity operator

\begin{tcolorbox}[
	boxrule=0pt,
	sharp corners
	]
	\scriptsize
	\begin{lstlisting}
For[i = 1, i <= 2^L, i += 1,
pair[i] = Position[SPINQBasis, Reverse[SPINQBasis[[i]]]][[1, 1]]];
	\end{lstlisting}
\end{tcolorbox}

Using this basis one can determine the parity of any state  $|\psi\rangle$. One may expand the state in the following quantum spin basis
\begin{align}
	|\psi\rangle= \sum \mathcal{C}_i   |P_i\rangle,
\end{align}
with $|P_i\rangle$ being spin basis and $\mathcal{C}_i=\langle P_i|\psi \rangle$. If we assume that the parity of
the state $|\psi\rangle$ is even, it is possible to rewrite the expansion as follows
\begin{align}
	|\psi\rangle= \sum \frac12 \left(\mathcal{C}_i +\mathcal{C}_j \right)   |P_i\rangle,
\end{align}
which $C_j$ is projection of $\psi$ on the basis  $|P_j>$ that  is pair to $|P_i>$ with parity operator
$C_j=\langle P_i|\Pi|\psi \rangle=\langle P_j|\psi \rangle$.
To ascertain the parity of the state, we employ the probability amplitude. Should the probability amplitude equal one, it indicates that our assumption regarding the state's parity is correct, leading us to deduce that the initial state possesses even parity. Conversely, if the probability amplitude differs from one, then the state's parity is deemed odd. This condition can be represented as follows:

\begin{tcolorbox}[
	boxrule=0pt,
	sharp corners
	]
	\scriptsize
	\begin{lstlisting}
	eigensystem = Eigensystem[N[H]];
	sortedEigensystem = Transpose[SortBy[Transpose[eigensystem], First]];
	For[i = 1, i <= 2^L, i += 1,
	If[
	0.8<Sum[1/4(sortedEigensystem[[2, i, j]]+sortedEigensystem[[2, i, pair[j]]])^2,
	{j, 1, 2^L}]<1.2,parity[i] = 1, parity[i] = -1]]
	\end{lstlisting}
\end{tcolorbox}

At this point, we are able to categorize the eigenvalues into two groups: those associated with odd parity eigenvectors and those with even parity eigenvectors

\begin{tcolorbox}[
	boxrule=0pt,
	sharp corners
	]
	\scriptsize
	\begin{lstlisting}
	EigenValueEven = {};
	EigenValueOdd = {};
	For[i = 1, i <= 2^L, i += 1,
	If[parity[i] == 1, AppendTo[EigenValueEven, sortedEigensystem[[1, i]]], 
	AppendTo[EigenValueOdd, sortedEigensystem[[1, i]]]]]
	\end{lstlisting}
\end{tcolorbox}

After dividing the eigenvalues into two sectors—negative and positive—based on parity symmetry, we can normalize the eigenvalues within each sector as detailed in Appendix A. For the positive sector, we normalize the level spacing as follows

\begin{tcolorbox}[
	boxrule=0pt,
	sharp corners
	]
	\scriptsize
	\begin{lstlisting}
	leveleven = {};
	Delta = Floor[Sqrt[Length[EigenValueEven]]];
	For[i = 10 Delta, i <= Length[EigenValueEven] - 10 Delta,i += 1,
	AppendTo[leveleven,
	1/(EigenValueEven[[i + Delta]] - EigenValueEven[[i - Delta]]) (EigenValueEven[[i + 1]]
		- EigenValueEven[[i]]) ]];
	\end{lstlisting}
\end{tcolorbox}

\begin{tcolorbox}[
	boxrule=0pt,
	sharp corners
	]
	\scriptsize
	\begin{lstlisting}
	EigenValueEvenUnfold={};
	For[m = 1, m <= Length[leveleven], m += 1,
		AppendTo[EigenValueEvenUnfold, (leveleven[[m]])/Mean[leveleven]]];
	\end{lstlisting}
\end{tcolorbox}
We can do the same for the odd  parity  sector

\begin{tcolorbox}[
	boxrule=0pt,
	sharp corners
	]
	\scriptsize
	\begin{lstlisting}
	levelodd = {};
	Delta = Floor[Sqrt[Length[EigenValueOdd]]];
	For[i = 10 Delta, i <= Length[EigenValueOdd] - 10 Delta, i +=1,
	AppendTo[levelodd, 
	1/(EigenValueOdd[[i + Delta]]-EigenValueOdd[[i - Delta]]) (EigenValueOdd[[i + 1]] 
		- EigenValueOdd[[i]])]]
	\end{lstlisting}
\end{tcolorbox}

\begin{tcolorbox}[
	boxrule=0pt,
	sharp corners
	]
	\scriptsize
	\begin{lstlisting}
	EigenValueOddUnfold={};
	For[m = 1, m <= Length[levelodd], m += 1,
	AppendTo[EigenValueOddUnfold, (levelodd[[m]])/Mean[levelodd]]
		];
	\end{lstlisting}
\end{tcolorbox}

Finally, we can plot the histogram of level spacing

\begin{tcolorbox}[
	boxrule=0pt,
	sharp corners
	]
	\scriptsize
	\begin{lstlisting}
	Histogram[EigenValueEvenUnfold, {0.1}, "ProbabilityDensity", Frame -> True]
	Histogram[EigenValueOddUnfold, {0.1}, "ProbabilityDensity", Frame -> True]
	\end{lstlisting}
\end{tcolorbox}

\subsection*{Matrix elements of an Operator in energy eigenstates}

We should first select an operator for which we
would like to compute its matrix elements. The  operator we have considered is $S_x$ 
\begin{tcolorbox}[
	boxrule=0pt,
	sharp corners
	]
	\scriptsize
	\begin{lstlisting}[mathescape=true]
Sx[L_]=Sum[sigmaz[i, L], {i, 1, L}];
	\end{lstlisting}
\end{tcolorbox}
We also need to find the eigenvectors of the Hamiltonian in the even parity (or odd; we choose the even sector here). We use the same algorithm used previously to find the eigenvalues.
\begin{tcolorbox}[
	boxrule=0pt,
	sharp corners
	]
	\scriptsize
	\begin{lstlisting}[mathescape=true]
eigensystem = Eigensystem[N[H]];
sortedEigensystem = Transpose[SortBy[Transpose[eigensystem], First]];
Upstate = Table[1, {i, 1, L}];
Downstate = Table[0, {i, 1, L}];
SPINQBasis = Permutations[Join[Upstate, Downstate], {L}];
For[i = 1, i <= 2^L, i += 1, 
pair[i] = Position[SPINQBasis, Reverse[SPINQBasis[[i]]]][[1, 1]];];
For[i = 1, i <= 2^L, i += 1,
project = 0;
For[j = 1, j <= 2^L, j += 1,
project = 
project + 
1/4 (sortedEigensystem[[2, i, j]] + 
sortedEigensystem[[2, i, pair[j]]])^2;];
If[0.8 < project < 1.2, parity[i] = 1, parity[i] = -1]

];
EigenVectorEven = {};
EigenVectorOdd = {};
For[i = 1, i <= 2^L, i += 1,
If[parity[i] == 1, 
AppendTo[EigenVectorEven, sortedEigensystem[[2, i]]], 
AppendTo[EigenVectorEven, sortedEigensystem[[2, i]]]]];
	\end{lstlisting}
\end{tcolorbox}

The matrix elements of the operator $S_x$ in energy eigenstates can be then computed as follows
\begin{tcolorbox}[
	boxrule=0pt,
	sharp corners
	]
	\scriptsize
	\begin{lstlisting}[mathescape=true]
OinEVecH = 
ParallelTable[
ConjugateTranspose[EigenVectorEven[[m]]] . Sx[L] . 
EigenVectorEven[[n]], {m, 1, Length[EigenVectorEven]}, {n, 1, 
	Length[EigenVectorEven]}];
	\end{lstlisting}
\end{tcolorbox}
 One can  plot the absolute value of matrix elements 
 using the following command ( see Figure 1) 
\begin{tcolorbox}[
	boxrule=0pt,
	sharp corners
	]
	\scriptsize
	\begin{lstlisting}[mathescape=true]
MatrixPlot[Abs[OinEVecH]]
	\end{lstlisting}
\end{tcolorbox}
\subsection*{Expectation Value of an Operator}
As in the previous example, we have considered
the operator $S_x$ 
\begin{tcolorbox}[
	boxrule=0pt,
	sharp corners
	]
	\scriptsize
	\begin{lstlisting}[mathescape=true]
Sx[L_]=Sum[sigmax[i, L], {i, 1, L}];
	\end{lstlisting}
\end{tcolorbox}
To compute the expectation values of the operator, we require the eigenvalues and corresponding normalized eigenvectors of the Hamiltonian restricted to one symmetry sector. These quantities were obtained in the previous example and are reused here. Furthermore, we must evaluate the representation of the operator in the energy eigenbasis of this symmetry sector
\begin{tcolorbox}[
	boxrule=0pt,
	sharp corners
	]
	\scriptsize
	\begin{lstlisting}[mathescape=true]
OinEVecH = 
ParallelTable[
N[Dot[ConjugateTranspose[EigenVectorEven[[n]]] . Sx[L] . 
EigenVectorEven[[m]]]], {m, 1, Length[EigenVectorEven]}, {n, 1, 
	Length[EigenVectorEven]}];
	\end{lstlisting}
\end{tcolorbox}
Now we have all the ingredients   to compute the expectation value of $ S_x $ for any specified values of $\theta $ and $\phi $ (initial state) at any time $t$

\begin{tcolorbox}[
	boxrule=0pt,
	sharp corners
	]
	\scriptsize
	\begin{lstlisting}[mathescape=true]
Do[Do[
      listEV[theta, phi] = {};
        listc = {};
       For[m = 1, m <= Length[EigenValueEven], m += 1, 
AppendTo[listc, {m, N[Dot[ConjugateTranspose[EigenVectorEven[[m]]] . initialState[theta, phi, L]]]}]];
                 
 EVO[t_] := 
       ParallelSum[
         E^(I (EigenValueEven[[m]] - EigenValueEven[[n]]) t)*    Conjugate[listc[[n, 2]]]*
            listc[[m, 2]]  *OinEVecH[[m, n]], {n, 1, Length[EigenValueEven]}, {m, 
	              1, Length[EigenValueEven]}] ;
           
For[t = 0, t <= 30, t += 2/10,
     AppendTo[listEV[theta, phi], {t, EVO[t]}]], {phi, {Pi/2}}], {theta, {Pi/2}}]
	\end{lstlisting}
\end{tcolorbox}
In this code, we have set  $\theta = \frac{\pi}{2}$ and $ \phi = \frac{\pi}{2} $

\begin{tcolorbox}[
	boxrule=0pt,
	sharp corners
	]
	\scriptsize
	\begin{lstlisting}[mathescape=true]
ListLinePlot[{listEV[Pi/2, Pi/2], {{0, 0}, {30, 0}}}, 
	FrameTicks -> {{Automatic, None}}, 
	PlotStyle -> {{Blue}, {Orange, Dashed}},
	PlotRange -> All,
	Frame -> True,
	FrameLabel -> {"t", "< \!\(\*SubscriptBox[\(S\), \(x\)]\) >"},
	LabelStyle -> Directive[FontSize -> 10], RotateLabel -> True, 
	FrameStyle -> Directive[Black, Bold] 
]
	\end{lstlisting}
\end{tcolorbox}

\subsection*{Absolute Value of Effective Inverse Temperature}

To compute the effective inverse temperature, we need the eigenvalues of one symmetry sector of the Hamiltonian, and a list of these eigenvalues that appear in the product for a general inverse temperature  $\beta$
\begin{tcolorbox}[
	boxrule=0pt,
	sharp corners
	]
	\scriptsize
	\begin{lstlisting}[mathescape=true]
betaEigenValueEven =-beta*EigenValueEven;
	\end{lstlisting}
\end{tcolorbox}

Using these lists, we obtain the trace of the canonical ensemble for a general inverse temperature $\beta$. Finally, by numerically solving the equation
$
\operatorname{Tr}(\rho_{\mathrm{th}} H) - \operatorname{Tr}(\rho_0 H) = 0,
$
we find the absolute value of the effective inverse temperature.

\begin{tcolorbox}[
	boxrule=0pt,
	sharp corners
	]
	\scriptsize
	\begin{lstlisting}[mathescape=true]
Rho[theta_, phi_, L_] := 
Partition[initialState[theta, phi, L], 1] . 
Transpose[Partition[initialState[theta, phi, L], 1]];
list = {};
For[beta = -2, beta <= 2, beta += 1/10,
x = ParallelSum[
EigenValueEven[[i]] E^betaEigenValueEven[[i]], {i, 1, 
	Length[EigenValueEven]}]/
ParallelSum[E^
betaEigenValueEven[[i]], {i, 1, Length[betaEigenValueEven]}];

AppendTo[list, {beta, x - Tr[Dot[N[H], N[Rho[Pi/2, Pi/2, L]]]]}];
];
f = Interpolation[Re[list]];
beta = FindRoot[f[y], {y, 0}][[1, 2]]
\end{lstlisting}
\end{tcolorbox}

\section{Reproducible Julia implementation}
\label{app:julia}

This Julia implementation follows the same conventions; its required packages are listed in the first lines below. A single Hermitian eigensystem keeps every ordered eigenvalue paired with the correct eigenvector.

\begin{tcolorbox}[breakable,boxrule=0pt,sharp corners]
\small
\begin{lstlisting}
using LinearAlgebra
using SparseArrays
using Statistics
using Roots
using Plots
using InteractiveUtils
using Pkg

versioninfo()
Pkg.status()

const id2 = Matrix{ComplexF64}(I, 2, 2)
const sigma_x = ComplexF64[0 1; 1 0]
const sigma_y = ComplexF64[0 -im; im 0]
const sigma_z = ComplexF64[1 0; 0 -1]

function sigma_at_n(mat, n, L)
    ops = [copy(id2) for _ in 1:L]
    ops[n] = ComplexF64.(mat)
    return foldl(kron, ops)
end

function two_site_at_n(mat1, i, mat2, j, L)
    ops = [copy(id2) for _ in 1:L]
    ops[i] = ComplexF64.(mat1)
    ops[j] = ComplexF64.(mat2)
    return foldl(kron, ops)
end

sigmax(n, L) = sigma_at_n(sigma_x, n, L)
sigmay(n, L) = sigma_at_n(sigma_y, n, L)
sigmaz(n, L) = sigma_at_n(sigma_z, n, L)
zz(n, L) = two_site_at_n(sigma_z, n, sigma_z, n+1, L)

function H(L, g, h; J=1.0)
    d = 2^L
    out = zeros(ComplexF64, d, d)
    for i in 1:L
        out .-= g .* sigmax(i, L)
        out .-= h .* sigmaz(i, L)
    end
    for i in 1:(L-1)
        out .-= J .* zz(i, L)
    end
    return out
end

Sx(L) = sum(sigmax(i, L) for i in 1:L)
Sz(L) = sum(sigmaz(i, L) for i in 1:L)

function initial_state(theta, phi, L)
    v = ComplexF64[
        cos(theta/2),
        exp(im*phi) * sin(theta/2)
    ]
    return foldl(kron, [copy(v) for _ in 1:L])
end

function reverse_bits(x::Int, L::Int)
    y = 0
    for _ in 1:L
        y = (y << 1) | (x & 1)
        x >>= 1
    end
    return y
end

function reflection_operator(L)
    d = 2^L
    rows = [reverse_bits(j-1, L) + 1 for j in 1:d]
    cols = collect(1:d)
    return sparse(rows, cols, ones(ComplexF64, d), d, d)
end

function spinflip_operator(L)
    d = 2^L
    rows = reverse(collect(1:d))
    cols = collect(1:d)
    return sparse(rows, cols, ones(ComplexF64, d), d, d)
end

function symmetry_basis(L, qR; qF=nothing)
    qR in (-1, 1) || error("qR must be +1 or -1")
    qF === nothing ||
        (qF in (-1, 1) || error("qF must be +1 or -1"))

    d = 2^L
    id = Matrix{ComplexF64}(I, d, d)
    reflection = Matrix(reflection_operator(L))
    projector = (id + qR .* reflection) ./ 2

    if qF !== nothing
        flip = Matrix(spinflip_operator(L))
        projector = projector * (id + qF .* flip) ./ 2
    end

    proj_eig = eigen(Hermitian(projector))
    keep = findall(abs.(proj_eig.values .- 1) .< 1e-10)
    return proj_eig.vectors[:, keep]
end

function sector_data(L, g, h; J=1.0, qR=1, qF=nothing)
    basis = symmetry_basis(L, qR; qF=qF)
    hfull = H(L, g, h; J=J)
    hsector = basis' * hfull * basis
    osector = basis' * Sx(L) * basis

    eigsys = eigen(Hermitian(hsector))
    order = sortperm(eigsys.values)
    energies = real.(eigsys.values[order])
    U = eigsys.vectors[:, order]
    oenergy = U' * osector * U

    return (
        L=L, basis=basis, hfull=hfull,
        energies=energies, U=U, oenergy=oenergy
    )
end
\end{lstlisting}
\end{tcolorbox}

\begin{tcolorbox}[breakable,boxrule=0pt,sharp corners]
\small
\begin{lstlisting}
state_in_energy(data, psi) = data.U' * (data.basis' * psi)

function expectation_curve(data, psi, times)
    coeff = state_in_energy(data, psi)
    return [
        begin
            a = coeff .* exp.(-im .* data.energies .* t)
            (t, real(dot(a, data.oenergy * a)))
        end
        for t in times
    ]
end

function thermal_energy(beta, energies)
    x = -beta .* energies
    x .-= maximum(x)
    weights = exp.(x)
    return dot(energies, weights) / sum(weights)
end

function effective_beta(target, energies)
    minimum(energies) < target < maximum(energies) ||
        error("A finite beta does not exist for this target.")

    f(beta) = thermal_energy(beta, energies) - target
    bound = 1.0
    while f(-bound) * f(bound) > 0 && bound < 2.0^20
        bound *= 2
    end
    f(-bound) * f(bound) <= 0 ||
        error("Could not bracket beta.")
    return find_zero(f, (-bound, bound), Bisection())
end

function canonical_mean(beta, energies, diagonal)
    x = -beta .* energies
    x .-= maximum(x)
    weights = exp.(x)
    return dot(weights, diagonal) / sum(weights)
end

function weighted_thermal_energy(beta, sector_energies,
                                 sector_weights)
    return sum(
        w * thermal_energy(beta, e)
        for (w, e) in zip(sector_weights, sector_energies)
    )
end

function effective_beta_weighted(target, sector_energies,
                                 sector_weights)
    low = sum(
        w * minimum(e)
        for (w, e) in zip(sector_weights, sector_energies)
    )
    high = sum(
        w * maximum(e)
        for (w, e) in zip(sector_weights, sector_energies)
    )
    low < target < high ||
        error("A finite constrained beta does not exist.")

    f(beta) = weighted_thermal_energy(
        beta, sector_energies, sector_weights
    ) - target
    bound = 1.0
    while f(-bound) * f(bound) > 0 && bound < 2.0^20
        bound *= 2
    end
    f(-bound) * f(bound) <= 0 ||
        error("Could not bracket constrained beta.")
    return find_zero(f, (-bound, bound), Bisection())
end

function weighted_canonical_mean(beta, sector_energies,
                                 sector_diagonals,
                                 sector_weights)
    return sum(
        w * canonical_mean(beta, e, d)
        for (w, e, d) in zip(
            sector_weights, sector_energies, sector_diagonals
        )
    )
end

Ldyn = 9
g = -1.05
psiY = initial_state(pi/2, pi/2, Ldyn)
data_chaotic = sector_data(Ldyn, g, 0.5; J=1.0, qR=1)
data_integrable_dynamics = sector_data(
    Ldyn, g, 0.0; J=1.0, qR=1
)
data_integrable_plus = sector_data(
    Ldyn, g, 0.0; J=1.0, qR=1, qF=1
)
data_integrable_minus = sector_data(
    Ldyn, g, 0.0; J=1.0, qR=1, qF=-1
)

curve_chaotic = expectation_curve(
    data_chaotic, psiY, 0:0.2:30
)
curve_integrable = expectation_curve(
    data_integrable_dynamics, psiY, 0:0.2:30
)

target_energy = real(dot(
    psiY, data_chaotic.hfull * psiY
))
beta_eff = effective_beta(
    target_energy, data_chaotic.energies
)
thermal_Sx = canonical_mean(
    beta_eff,
    data_chaotic.energies,
    real.(diag(data_chaotic.oenergy))
)

target_energy_integrable = real(dot(
    psiY,
    data_integrable_dynamics.hfull * psiY
))
flip = Matrix(spinflip_operator(Ldyn))
id_full = Matrix{ComplexF64}(I, 2^Ldyn, 2^Ldyn)
pF_plus = real(dot(psiY, ((id_full + flip) / 2) * psiY))
pF_minus = 1 - pF_plus
sector_weights_integrable = [pF_plus, pF_minus]
sector_energies_integrable = [
    data_integrable_plus.energies,
    data_integrable_minus.energies
]
sector_diagonals_integrable = [
    real.(diag(data_integrable_plus.oenergy)),
    real.(diag(data_integrable_minus.oenergy))
]

beta_eff_integrable = effective_beta_weighted(
    target_energy_integrable,
    sector_energies_integrable,
    sector_weights_integrable
)
thermal_Sx_integrable = weighted_canonical_mean(
    beta_eff_integrable,
    sector_energies_integrable,
    sector_diagonals_integrable,
    sector_weights_integrable
)

p_chaotic = plot(
    first.(curve_chaotic), last.(curve_chaotic),
    xlabel="t", ylabel="<S_x(t)>", linewidth=2,
    label="dynamics"
)
hline!(p_chaotic, [thermal_Sx],
       linestyle=:dash, color=:black, label="thermal reference")

p_integrable = plot(
    first.(curve_integrable), last.(curve_integrable),
    xlabel="t", ylabel="<S_x(t)>", linewidth=2,
    label="dynamics"
)
hline!(p_integrable, [thermal_Sx_integrable],
       linestyle=:dash, color=:black,
       label="weighted-sector reference")

plot(p_chaotic, p_integrable, layout=(1, 2), size=(900, 350))
\end{lstlisting}
\end{tcolorbox}

\begin{tcolorbox}[breakable,boxrule=0pt,sharp corners]
\small
\begin{lstlisting}
function unfolded_spacings(energies)
    e = sort(real.(energies))
    n = length(e)
    window = floor(Int, sqrt(n))
    n > 2*window + 1 || error("Symmetry sector is too small.")
    scale = 1e-12 * max(1.0, maximum(abs.(e)))

    raw = [
        2*window*(e[i+1] - e[i]) /
        (e[i+window] - e[i-window])
        for i in (window+1):(n-window)
        if e[i+window] - e[i-window] > scale
    ]
    isempty(raw) && error("No usable unfolded spacings.")
    return raw ./ mean(raw)
end

function gap_ratios(energies)
    e = sort(real.(energies))
    gaps = diff(e)
    scale = 1e-12 * max(1.0, maximum(abs.(e)))
    return [
        min(a, b) / max(a, b)
        for (a, b) in zip(gaps[1:end-1], gaps[2:end])
        if min(a, b) > scale
    ]
end

# Mixed field: reflection-even sector.
level_chaotic = sector_data(
    10, -1.05, 0.5; J=1.0, qR=1
)

# h=0: resolve reflection and global spin flip.
level_integrable = sector_data(
    10, -1.05, 0.0; J=1.0, qR=1, qF=1
)

s_chaotic = unfolded_spacings(level_chaotic.energies)
s_integrable = unfolded_spacings(level_integrable.energies)
r_chaotic = gap_ratios(level_chaotic.energies)
r_integrable = gap_ratios(level_integrable.energies)

println((
    length(level_chaotic.energies),
    length(level_integrable.energies),
    mean(r_chaotic),
    mean(r_integrable)
))

histogram(
    [s_chaotic, s_integrable],
    bins=0:0.1:4, normalize=:pdf,
    labels=["mixed field" "integrable"],
    xlabel="unfolded spacing", ylabel="density"
)

heatmap(abs.(data_chaotic.oenergy),
        xlabel="m", ylabel="n",
        title="symmetry-resolved |<E_n|S_x|E_m>|")
\end{lstlisting}
\end{tcolorbox}

For either language, a fresh session should be used and the package/kernel versions, system size, couplings, sector charges, ensemble convention, and plotting bins should be recorded with the exported data. The two-dimensional scans used for Figures~6--7 evaluate Eq.~\eqref{EE} and the bracketed effective-temperature routine on the declared $(\theta,\phi)$ grid; the rendering commands are not repeated in these compact listings.

\bibliography{literature}

@article{Von,
  author = {J. von Neumann},
  title = {Quantum Mechanics and the Foundations of Statistical Mechanics},
  journal = {Zeitschrift f\"ur Physik},
  volume = {57},
  year = {1929},
  pages = {30--70},
  note = {Translated to English in Eur. Phys. J. H 35, 201 (2010)}
}

@article{Kinoshita:2006,
  author = {T. Kinoshita and T. Wenger and D. S. Weiss},
  title = {A quantum Newton's cradle},
  journal = {Nature},
  volume = {440},
  year = {2006},
  pages = {900--903},
  doi = {10.1038/nature04693}
}

@article{Hofferberth:2007,
  author = {S. Hofferberth and I. Lesanovsky and B. Fischer and T. Schumm and J. Schmiedmayer},
  title = {Non-equilibrium coherence dynamics in one-dimensional Bose gases},
  journal = {Nature},
  volume = {449},
  year = {2007},
  pages = {324--327},
  doi = {10.1038/nature06149}
}

@article{Trotzky:2012,
  author = {S. Trotzky and Y.-A. Chen and A. Flesch and I. P. McCulloch and U. Schollw{\"o}ck and J. Eisert and I. Bloch},
  title = {Probing the relaxation towards equilibrium in an isolated strongly correlated one-dimensional Bose gas},
  journal = {Nature Phys.},
  volume = {8},
  year = {2012},
  pages = {325--330},
  doi = {10.1038/nphys2232}
}

@article{Smith:2016,
  author = {J. Smith and A. Lee and P. Richerme and B. Neyenhuis and P. W. Hess and P. Hauke and M. Heyl and D. A. Huse and C. Monroe},
  title = {Many-body localization in a quantum simulator with programmable random disorder},
  journal = {Nature Phys.},
  volume = {12},
  year = {2016},
  pages = {907--911},
  doi = {10.1038/nphys3783}
}

@article{Zhang:2017,
  author = {J. Zhang and G. Pagliani and P. W. Hess and A. Kyprianidis and P. Becker and H. Kaplan and A. V. Gorshkov and Z.-X. Gong and C. Monroe},
  title = {Observation of a many-body dynamical phase transition with a 53-qubit quantum simulator},
  journal = {Nature},
  volume = {551},
  year = {2017},
  pages = {601--604},
  doi = {10.1038/nature24654}
}

@article{Neyenhuis:2017,
  author = {B. Neyenhuis and J. Zhang and P. W. Hess and J. Smith and A. Lee and P. Richerme and Z.-X. Gong and A. V. Gorshkov and C. Monroe},
  title = {Observation of prethermalization in long-range interacting spin chains},
  journal = {Sci. Adv.},
  volume = {3},
  year = {2017},
  pages = {e1700672},
  doi = {10.1126/sciadv.1700672}
}

@article{Yukalov:2011,
  author = {V. I. Yukalov},
  title = {Equilibration and thermalization in finite quantum systems},
  journal = {Laser Phys.},
  volume = {21},
  year = {2011},
  pages = {1--46},
  doi = {10.1134/S1054660X1103019X}
}

@article{Polkovnikov:2011,
  author = {A. Polkovnikov and K. Sengupta and A. Silva and M. Vengalattore},
  title = {Nonequilibrium dynamics of closed interacting quantum systems},
  journal = {Rev. Mod. Phys.},
  volume = {83},
  year = {2011},
  pages = {863},
  doi = {10.1103/RevModPhys.83.863}
}

@article{Eisert:2015,
  author = {J. Eisert and M. Friesdorf and C. Gogolin},
  title = {Quantum many-body systems out of equilibrium},
  journal = {Nature Phys.},
  volume = {11},
  year = {2015},
  pages = {124--130},
  doi = {10.1038/nphys3215}
}

@article{Ueda:2018,
  author = {M. Ueda},
  title = {Quantum equilibration, thermalization and prethermalization in ultracold atoms},
  journal = {Nature Rev. Phys.},
  volume = {1},
  year = {2019},
  pages = {197--207},
  doi = {10.1038/s42254-018-0008-5}
}

@article{Altman:2015,
  author = {E. Altman},
  title = {Non-equilibrium quantum dynamics in ultra-cold quantum gases},
  journal = {Lecture Notes of the Les Houches Summer School},
  year = {2015},
  note = {arXiv:1512.00870 [cond-mat.quant-gas]},
  archiveprefix = {arXiv},
  eprint = {1512.00870},
  primaryclass = {cond-mat.quant-gas}
}

@article{Borgonovi:2016,
  author = {F. Borgonovi and F. M. Izrailev and L. F. Santos and V. G. Zelevinsky},
  title = {Quantum chaos and thermalization in isolated systems of interacting particles},
  journal = {Phys. Rep.},
  volume = {626},
  year = {2016},
  pages = {1--58},
  doi = {10.1016/j.physrep.2016.02.005}
}

@article{Deutsch:1991,
  author = {J. M. Deutsch},
  title = {Quantum statistical mechanics in a closed system},
  journal = {Phys. Rev. A},
  volume = {43},
  year = {1991},
  pages = {2046},
  doi = {10.1103/PhysRevA.43.2046}
}

@article{Srednicki:1994mfb,
  author = {M. Srednicki},
  title = {Chaos and Quantum Thermalization},
  journal = {Phys. Rev. E},
  volume = {50},
  year = {1994},
  pages = {888},
  doi = {10.1103/PhysRevE.50.888},
  archiveprefix = {arXiv},
  eprint = {cond-mat/9403051},
  primaryclass = {cond-mat}
}

@article{Serdnicki:1999,
  author = {M. Srednicki},
  title = {The approach to thermal equilibrium in quantized chaotic systems},
  journal = {J. Phys. A},
  volume = {32},
  year = {1999},
  pages = {1163}
}

@article{Bohigas:1983er,
  author = {O. Bohigas and M. J. Giannoni and C. Schmit},
  title = {Characterization of chaotic quantum spectra and universality of level fluctuation laws},
  journal = {Phys. Rev. Lett.},
  volume = {52},
  year = {1984},
  pages = {1--4},
  doi = {10.1103/PhysRevLett.52.1}
}

@article{DAlessio:2015qtq,
  author = {L. D'Alessio and Y. Kafri and A. Polkovnikov and M. Rigol},
  title = {From quantum chaos and eigenstate thermalization to statistical mechanics and thermodynamics},
  journal = {Adv. Phys.},
  volume = {65},
  year = {2016},
  number = {3},
  pages = {239--362},
  doi = {10.1080/00018732.2016.1198134},
  archiveprefix = {arXiv},
  eprint = {1509.06411},
  primaryclass = {cond-mat.stat-mech}
}

@article{Deutsch:2018,
  author = {J. M. Deutsch},
  title = {Eigenstate thermalization hypothesis},
  journal = {Rep. Prog. Phys.},
  volume = {81},
  year = {2018},
  number = {8},
  pages = {082001},
  doi = {10.1088/1361-6633/aac9f1}
}

@article{Mori:2017qhg,
  author = {T. Mori and T. N. Ikeda and E. Kaminishi and M. Ueda},
  title = {Thermalization and prethermalization in isolated quantum systems: A theoretical overview},
  journal = {J. Phys. B},
  volume = {51},
  year = {2018},
  number = {11},
  pages = {112001},
  doi = {10.1088/1361-6455/aabcdf},
  archiveprefix = {arXiv},
  eprint = {1712.08790},
  primaryclass = {cond-mat.stat-mech}
}

@misc{Pappalardi,
  author = {S. Pappalardi},
  title = {Lecture notes on quantum thermalization},
  howpublished = {\url{https://www.qhaos.org/teaching}}
}

@article{Short:2011pvc,
  author = {A. J. Short and T. C. Farrelly},
  title = {Quantum equilibration in finite time},
  journal = {New J. Phys.},
  volume = {14},
  year = {2012},
  number = {1},
  pages = {013063},
  doi = {10.1088/1367-2630/14/1/013063},
  archiveprefix = {arXiv},
  eprint = {1110.5759},
  primaryclass = {quant-ph}
}

@article{Rigol:2007,
  author = {M. Rigol and V. Dunjko and V. Yurovsky and M. Olshanii},
  title = {Relaxation in a completely integrable many-body quantum system: An ab initio study of the dynamics of the highly excited states of 1D lattice hard-core bosons},
  journal = {Phys. Rev. Lett.},
  volume = {98},
  year = {2007},
  pages = {050405},
  doi = {10.1103/PhysRevLett.98.050405},
  archiveprefix = {arXiv},
  eprint = {cond-mat/0604476},
  primaryclass = {cond-mat.stat-mech}
}

@book{Book:2009,
  author = {M. Cencini and F. Cecconi and A. Vulpiani},
  title = {Chaos: From Simple Models to Complex Systems},
  publisher = {World Scientific},
  year = {2009}
}

@article{Maldacena:2015waa,
  author = {J. Maldacena and S. H. Shenker and D. Stanford},
  title = {A bound on chaos},
  journal = {JHEP},
  volume = {08},
  year = {2016},
  pages = {106},
  doi = {10.1007/JHEP08(2016)106},
  archiveprefix = {arXiv},
  eprint = {1503.01409},
  primaryclass = {hep-th}
}

@article{Hashimoto:2020xfr,
  author = {K. Hashimoto and K. B. Huh and K. Y. Kim and R. Watanabe},
  title = {Exponential growth of out-of-time-order correlator without chaos: Inverted harmonic oscillator},
  journal = {JHEP},
  volume = {11},
  year = {2020},
  pages = {068},
  doi = {10.1007/JHEP11(2020)068},
  archiveprefix = {arXiv},
  eprint = {2007.04746},
  primaryclass = {hep-th}
}

@article{Wigner:1955,
  author = {E. Wigner},
  title = {Characteristic Vectors of Bordered Matrices with Infinite Dimensions},
  journal = {Ann. Math.},
  volume = {62},
  year = {1955},
  pages = {548--564}
}

@article{Wigner:1957,
  author = {E. Wigner},
  title = {Characteristics Vectors of Bordered Matrices with Infinite Dimensions {II}},
  journal = {Ann. Math.},
  volume = {65},
  year = {1957},
  pages = {203--207}
}

@article{Wigner:1958,
  author = {E. Wigner},
  title = {On the Distribution of the Roots of Certain Symmetric Matrices},
  journal = {Ann. Math.},
  volume = {67},
  year = {1958},
  pages = {325--327}
}

@book{Mehta:2004,
  author = {M. L. Mehta},
  title = {Random Matrices},
  edition = {3rd},
  publisher = {Elsevier/Academic Press},
  address = {Amsterdam},
  year = {2004}
}

@article{Livan:2017,
  author = {G. Livan and M. Novaes and P. Vivo},
  title = {Introduction to Random Matrices: Theory and Practice},
  year = {2017},
  archiveprefix = {arXiv},
  eprint = {1712.07903},
  primaryclass = {math-ph}
}

@article{Atas:2013,
  author = {Y. Y. Atas and E. Bogomolny and O. Giraud and G. Roux},
  title = {Distribution of the ratio of consecutive level spacings in random matrix ensembles},
  journal = {Phys. Rev. Lett.},
  volume = {110},
  year = {2013},
  pages = {084101},
  doi = {10.1103/PhysRevLett.110.084101}
}

@article{Oganesyan:2007,
  author = {V. Oganesyan and D. A. Huse},
  title = {Localization of interacting fermions at high temperature},
  journal = {Phys. Rev. B},
  volume = {75},
  year = {2007},
  pages = {155111},
  doi = {10.1103/PhysRevB.75.155111}
}

@article{Banuls:2011vuw,
  author = {M. C. Ba{\~n}uls and J. I. Cirac and M. B. Hastings},
  title = {Strong and weak thermalization of infinite nonintegrable quantum systems},
  journal = {Phys. Rev. Lett.},
  volume = {106},
  year = {2011},
  number = {5},
  pages = {050405},
  doi = {10.1103/PhysRevLett.106.050405}
}

@article{Sun:2020ybj,
  author = {Z. H. Sun and J. Cui and H. Fan},
  title = {Quantum information scrambling in the presence of weak and strong thermalization},
  journal = {Phys. Rev. A},
  volume = {104},
  year = {2021},
  number = {2},
  pages = {022405},
  doi = {10.1103/PhysRevA.104.022405},
  archiveprefix = {arXiv},
  eprint = {2008.01477},
  primaryclass = {quant-ph}
}

@article{Chen:2021,
  author = {F. Chen and others},
  title = {Observation of strong and weak thermalization in a superconducting quantum processor},
  journal = {Phys. Rev. Lett.},
  volume = {127},
  year = {2021},
  pages = {020602},
  doi = {10.1103/PhysRevLett.127.020602}
}

@article{Lin:2016egw,
  author = {C. J. Lin and O. I. Motrunich},
  title = {Quasiparticle explanation of the weak-thermalization regime under quench in a nonintegrable quantum spin chain},
  journal = {Phys. Rev. A},
  volume = {95},
  year = {2017},
  number = {2},
  pages = {023621},
  doi = {10.1103/PhysRevA.95.023621},
  archiveprefix = {arXiv},
  eprint = {1610.04287},
  primaryclass = {cond-mat.stat-mech}
}

@article{Prazeres:2023hce,
  author = {L. F. dos Prazeres and T. R. de Oliveira},
  title = {Smooth crossover between weak and strong thermalization using rigorous bounds on equilibration of isolated systems},
  journal = {Physica A},
  volume = {653},
  pages = {130065},
  year = {2024},
  doi = {10.1016/j.physa.2024.130065},
  archiveprefix = {arXiv},
  eprint = {2310.13392},
  primaryclass = {quant-ph}
}

@article{Bhattacharjee:2022qjw,
  author = {B. Bhattacharjee and S. Sur and P. Nandy},
  title = {Probing quantum scars and weak ergodicity breaking through quantum complexity},
  journal = {Phys. Rev. B},
  volume = {106},
  year = {2022},
  number = {20},
  pages = {205150},
  doi = {10.1103/PhysRevB.106.205150},
  archiveprefix = {arXiv},
  eprint = {2208.05503},
  primaryclass = {quant-ph}
}

@article{Nandy:2023brt,
  author = {S. Nandy and B. Mukherjee and A. Bhattacharyya and A. Banerjee},
  title = {Quantum state complexity meets many-body scars},
  journal = {J. Phys. Condens. Matter},
  volume = {36},
  year = {2024},
  number = {15},
  pages = {155601},
  doi = {10.1088/1361-648X/ad1a7b},
  archiveprefix = {arXiv},
  eprint = {2305.13322},
  primaryclass = {quant-ph}
}

@article{Alishahiha:2024rwm,
  author = {M. Alishahiha and M. J. Vasli},
  title = {Thermalization in Krylov basis},
  journal = {Eur. Phys. J. C},
  volume = {85},
  year = {2025},
  number = {1},
  pages = {39},
  doi = {10.1140/epjc/s10052-025-13757-2},
  archiveprefix = {arXiv},
  eprint = {2403.06655},
  primaryclass = {quant-ph}
}

@article{Basko:2006,
  author = {D. M. Basko and I. L. Aleiner and B. L. Altshuler},
  title = {Metal-insulator transition in a weakly interacting many-electron system with localized single-particle states},
  journal = {Ann. Phys. (N.Y.)},
  volume = {321},
  year = {2006},
  pages = {1126--1205},
  doi = {10.1016/j.aop.2005.11.014}
}

@article{Serbyn:2013,
  author = {M. Serbyn and Z. Papi{\'c} and D. A. Abanin},
  title = {Local conservation laws and the structure of the many-body localized states},
  journal = {Phys. Rev. Lett.},
  volume = {111},
  year = {2013},
  pages = {127201},
  doi = {10.1103/PhysRevLett.111.127201}
}

@article{Huse:2014,
  author = {D. A. Huse and R. Nandkishore and V. Oganesyan},
  title = {Phenomenology of fully many-body-localized systems},
  journal = {Phys. Rev. B},
  volume = {90},
  year = {2014},
  pages = {174202},
  doi = {10.1103/PhysRevB.90.174202}
}

@article{Shiraishi:2017,
  author = {N. Shiraishi and T. Mori},
  title = {Systematic construction of counterexamples to the eigenstate thermalization hypothesis},
  journal = {Phys. Rev. Lett.},
  volume = {119},
  year = {2017},
  pages = {030601},
  doi = {10.1103/PhysRevLett.119.030601}
}

@article{Mori:2017,
  author = {T. Mori and N. Shiraishi},
  title = {Thermalization without eigenstate thermalization hypothesis after a quantum quench},
  journal = {Phys. Rev. E},
  volume = {96},
  year = {2017},
  pages = {022153},
  doi = {10.1103/PhysRevE.96.022153}
}

@article{Bernien:2017,
  author = {H. Bernien and S. Schwartz and A. Keesling and H. Levine and A. Omran and H. Pichler and S. Choi and A. S. Zibrov and M. Endres and M. Greiner and V. Vuleti{\'c} and M. D. Lukin},
  title = {Probing many-body dynamics on a 51-atom quantum simulator},
  journal = {Nature},
  volume = {551},
  year = {2017},
  pages = {579--584},
  doi = {10.1038/nature24654}
}

@article{Lesanovsky:2012,
  author = {I. Lesanovsky and H. Katsura},
  title = {Interacting Fibonacci anyons in a Rydberg gas},
  journal = {Phys. Rev. A},
  volume = {86},
  year = {2012},
  pages = {041601},
  doi = {10.1103/PhysRevA.86.041601}
}

@article{Turner:2018P,
  author = {C. J. Turner and A. A. Michailidis and D. A. Abanin and M. Serbyn and Z. Papi{\'c}},
  title = {Quantum scarred eigenstates in a Rydberg atom chain: Entanglement, breakdown of thermalization, and stability to perturbations},
  journal = {Phys. Rev. B},
  volume = {98},
  year = {2018},
  pages = {155134},
  doi = {10.1103/PhysRevB.98.155134}
}

@article{Turner:2018N,
  author = {C. J. Turner and A. A. Michailidis and D. A. Abanin and M. Serbyn and Z. Papi{\'c}},
  title = {Weak ergodicity breaking from quantum many-body scars},
  journal = {Nature Phys.},
  volume = {14},
  year = {2018},
  pages = {745--749},
  doi = {10.1038/s41567-018-0137-5}
}

@article{Heller:1984,
  author = {E. J. Heller},
  title = {Bound-state eigenfunctions of classically chaotic Hamiltonian systems: Scars of periodic orbits},
  journal = {Phys. Rev. Lett.},
  volume = {53},
  year = {1984},
  pages = {1515},
  doi = {10.1103/PhysRevLett.53.1515}
}

@article{Guo:2023wck,
  author = {Z. Guo and B. Liu and Y. Gao and A. Yang and J. Wang and J. Ma and L. Ying},
  title = {Origin of Hilbert-space quantum scars in unconstrained models},
  journal = {Phys. Rev. B},
  volume = {108},
  year = {2023},
  number = {7},
  pages = {075124},
  doi = {10.1103/PhysRevB.108.075124},
  archiveprefix = {arXiv},
  eprint = {2307.13297},
  primaryclass = {quant-ph}
}

@article{Khemani:2019vor,
  author = {V. Khemani and M. Hermele and R. Nandkishore},
  title = {Localization from Hilbert space shattering: From theory to physical realizations},
  journal = {Phys. Rev. B},
  volume = {101},
  year = {2020},
  number = {17},
  pages = {174204},
  doi = {10.1103/PhysRevB.101.174204},
  archiveprefix = {arXiv},
  eprint = {1904.04815},
  primaryclass = {cond-mat.stat-mech}
}

@article{Omran:2019unq,
  author = {A. Omran and H. Levine and A. Keesling and G. Semeghini and T. T. Wang and S. Ebadi and H. Bernien and A. S. Zibrov and H. Pichler and S. Choi and J. Cui and M. Rossignolo and P. Roushan and A. V. Gorshkov and M. D. Lukin},
  title = {Generation and manipulation of Schr{\"o}dinger cat states in Rydberg atom arrays},
  journal = {Science},
  volume = {365},
  year = {2019},
  number = {6453},
  pages = {aax9743},
  doi = {10.1126/science.aax9743}
}

@article{Serbyn:2020wys,
  author = {M. Serbyn and D. A. Abanin and Z. Papi{\'c}},
  title = {Quantum many-body scars and weak breaking of ergodicity},
  journal = {Nature Phys.},
  volume = {17},
  year = {2021},
  number = {6},
  pages = {675--685},
  doi = {10.1038/s41567-021-01230-2},
  archiveprefix = {arXiv},
  eprint = {2011.09486},
  primaryclass = {quant-ph}
}

@article{Maldacena:1997re,
  author = {J. M. Maldacena},
  title = {The large {N} limit of superconformal field theories and supergravity},
  journal = {Adv. Theor. Math. Phys.},
  volume = {2},
  year = {1998},
  pages = {231--252},
  doi = {10.4310/ATMP.1998.v2.n2.a1},
  archiveprefix = {arXiv},
  eprint = {hep-th/9711200}
}

@article{Aharony:1999ti,
  author = {O. Aharony and S. S. Gubser and J. M. Maldacena and H. Ooguri and Y. Oz},
  title = {Large {N} field theories, string theory and gravity},
  journal = {Phys. Rep.},
  volume = {323},
  year = {2000},
  pages = {183--386},
  doi = {10.1016/S0370-1573(99)00083-6},
  archiveprefix = {arXiv},
  eprint = {hep-th/9905111}
}

@article{Shenker:2013pqa,
  author = {S. H. Shenker and D. Stanford},
  title = {Black holes and the butterfly effect},
  journal = {JHEP},
  volume = {03},
  year = {2014},
  pages = {067},
  doi = {10.1007/JHEP03(2014)067},
  archiveprefix = {arXiv},
  eprint = {1306.0622},
  primaryclass = {hep-th}
}

@article{Shenker:2013yza,
  author = {S. H. Shenker and D. Stanford},
  title = {Multiple shocks},
  journal = {JHEP},
  volume = {12},
  year = {2014},
  pages = {046},
  doi = {10.1007/JHEP12(2014)046},
  archiveprefix = {arXiv},
  eprint = {1312.3296},
  primaryclass = {hep-th}
}

@article{Leichenauer:2014nxa,
  author = {S. Leichenauer},
  title = {Disrupting entanglement of black holes},
  journal = {Phys. Rev. D},
  volume = {90},
  year = {2014},
  number = {4},
  pages = {046009},
  doi = {10.1103/PhysRevD.90.046009},
  archiveprefix = {arXiv},
  eprint = {1405.7365},
  primaryclass = {hep-th}
}

@article{Roberts:2014isa,
  author = {D. A. Roberts and D. Stanford and L. Susskind},
  title = {Localized shocks},
  journal = {JHEP},
  volume = {03},
  year = {2015},
  pages = {051},
  doi = {10.1007/JHEP03(2015)051},
  archiveprefix = {arXiv},
  eprint = {1409.8180},
  primaryclass = {hep-th}
}

@misc{Kitaev:2015,
  author = {A. Kitaev},
  title = {A simple model of quantum holography},
  howpublished = {Talks at KITP Strings Seminar and Entanglement Program (Feb. 12, April 7, May 27, 2015)},
  year = {2015}
}

@article{Sachdev:2015efa,
  author = {S. Sachdev},
  title = {Bekenstein-Hawking entropy and strange metals},
  journal = {Phys. Rev. X},
  volume = {5},
  year = {2015},
  number = {4},
  pages = {041025},
  doi = {10.1103/PhysRevX.5.041025},
  archiveprefix = {arXiv},
  eprint = {1506.05111},
  primaryclass = {hep-th}
}

@article{Maldacena:2016hyu,
  author = {J. Maldacena and D. Stanford},
  title = {Remarks on the Sachdev-Ye-Kitaev model},
  journal = {Phys. Rev. D},
  volume = {94},
  year = {2016},
  number = {10},
  pages = {106002},
  doi = {10.1103/PhysRevD.94.106002},
  archiveprefix = {arXiv},
  eprint = {1604.07818},
  primaryclass = {hep-th}
}

@article{Jensen:2016pah,
  author = {K. Jensen},
  title = {Chaos in {AdS}$_2$ holography},
  journal = {Phys. Rev. Lett.},
  volume = {117},
  year = {2016},
  number = {11},
  pages = {111601},
  doi = {10.1103/PhysRevLett.117.111601},
  archiveprefix = {arXiv},
  eprint = {1605.06098},
  primaryclass = {hep-th}
}

@article{Maldacena:2016upp,
  author = {J. Maldacena and D. Stanford and Z. Yang},
  title = {Conformal symmetry and its breaking in two dimensional nearly anti-de-Sitter space},
  journal = {PTEP},
  volume = {2016},
  year = {2016},
  number = {12},
  pages = {12C104},
  doi = {10.1093/ptep/ptw124},
  archiveprefix = {arXiv},
  eprint = {1606.01857},
  primaryclass = {hep-th}
}

@article{Engelsoy:2016xyb,
  author = {J. Engels{\"o}y and T. G. Mertens and H. Verlinde},
  title = {An investigation of {AdS}$_2$ backreaction and holography},
  journal = {JHEP},
  volume = {07},
  year = {2016},
  pages = {139},
  doi = {10.1007/JHEP07(2016)139},
  archiveprefix = {arXiv},
  eprint = {1606.03438},
  primaryclass = {hep-th}
}

@article{Almheiri:2014cka,
  author = {A. Almheiri and J. Polchinski},
  title = {Models of {AdS}$_2$ backreaction and holography},
  journal = {JHEP},
  volume = {11},
  year = {2015},
  pages = {014},
  doi = {10.1007/JHEP11(2015)014},
  archiveprefix = {arXiv},
  eprint = {1402.6334},
  primaryclass = {hep-th}
}

@article{Saad:2019lba,
  author = {P. Saad and S. H. Shenker and D. Stanford},
  title = {{JT} gravity as a matrix integral},
  journal = {arXiv preprint arXiv:1903.11115},
  year = {2019},
  archiveprefix = {arXiv},
  eprint = {1903.11115},
  primaryclass = {hep-th}
}

@article{Sonner:2017hxc,
  author = {J. Sonner and M. Vielma},
  title = {Eigenstate thermalization in the Sachdev-Ye-Kitaev model},
  journal = {JHEP},
  volume = {11},
  year = {2017},
  pages = {149},
  doi = {10.1007/JHEP11(2017)149},
  archiveprefix = {arXiv},
  eprint = {1707.08013},
  primaryclass = {hep-th}
}

@article{Balasubramanian:2011ur,
  author = {V. Balasubramanian and A. Bernamonti and J. de Boer and N. Copland and B. Craps and E. Keski-Vakkuri and B. M{\"u}ller and A. Sch{\"a}fer and M. Shigemori and W. Staessens},
  title = {Holographic thermalization},
  journal = {Phys. Rev. D},
  volume = {84},
  year = {2011},
  pages = {026010},
  doi = {10.1103/PhysRevD.84.026010},
  archiveprefix = {arXiv},
  eprint = {1103.2683},
  primaryclass = {hep-th}
}

@article{Das:2010yw,
  author = {S. R. Das and T. Nishioka and T. Takayanagi},
  title = {Probe branes, time-dependent couplings and thermalization in AdS/CFT},
  journal = {JHEP},
  volume = {07},
  year = {2010},
  pages = {071},
  doi = {10.1007/JHEP07(2010)071},
  archiveprefix = {arXiv},
  eprint = {1005.3348},
  primaryclass = {hep-th}
}

@article{Bellantuono:2016mhl,
  author = {L. Bellantuono and R. A. Janik and J. Jalmuzna and P. Sur{\'o}wka and R. B. Peschanski},
  title = {Equilibration of a strongly interacting plasma: Holographic analysis of local and nonlocal probes},
  journal = {JHEP},
  volume = {12},
  year = {2016},
  pages = {128},
  doi = {10.1007/JHEP12(2016)128},
  archiveprefix = {arXiv},
  eprint = {1611.03326},
  primaryclass = {hep-th}
}

@article{Chen:2023iws,
  author = {Q. Chen and Z. Ning and Y. Tian and X. Wu and C.-Y. Zhang and H. Zhang},
  title = {Time evolution of Einstein-Maxwell-scalar black holes after a thermal quench},
  journal = {JHEP},
  volume = {10},
  year = {2023},
  pages = {176},
  doi = {10.1007/JHEP10(2023)176},
  archiveprefix = {arXiv},
  eprint = {2308.07666},
  primaryclass = {gr-qc}
}

@phdthesis{Fukushima:2025,
  author = {O. Fukushima},
  title = {Higher-form symmetry and eigenstate thermalization hypothesis},
  school = {Kyoto University},
  year = {2025},
  note = {Springer Theses, doi:10.1007/978-981-96-1643-5}
}

@article{Murthy:2023,
  author = {C. Murthy and A. Babakhani and F. Iniguez and M. Srednicki and N. Yunger Halpern},
  title = {Non-Abelian eigenstate thermalization hypothesis},
  journal = {Phys. Rev. Lett.},
  volume = {130},
  year = {2023},
  number = {14},
  pages = {140402},
  doi = {10.1103/PhysRevLett.130.140402}
}

@article{Sierant:2025,
  author = {P. Sierant and M. Lewenstein and A. Scardicchio and L. Vidmar and J. Zakrzewski},
  title = {Many-body localization in the age of classical computing},
  journal = {Rep. Prog. Phys.},
  volume = {88},
  year = {2025},
  number = {2},
  pages = {026502},
  doi = {10.1088/1361-6633/ad9756}
}

@article{Nandkishore:2015,
  author = {R. Nandkishore and D. A. Huse},
  title = {Many-body localization and thermalization in quantum statistical mechanics},
  journal = {Ann. Rev. Condens. Matter Phys.},
  volume = {6},
  year = {2015},
  pages = {15--38},
  doi = {10.1146/annurev-conmatphys-031214-014726},
  archiveprefix = {arXiv},
  eprint = {1404.0686},
  primaryclass = {cond-mat}
}

@article{Abanin:2019,
  author = {D. A. Abanin and E. Altman and I. Bloch and M. Serbyn},
  title = {Colloquium: Many-body localization, thermalization, and entanglement},
  journal = {Rev. Mod. Phys.},
  volume = {91},
  year = {2019},
  pages = {021001},
  doi = {10.1103/RevModPhys.91.021001},
  archiveprefix = {arXiv},
  eprint = {1804.11065},
  primaryclass = {cond-mat}
}

@article{Almeida:2026,
  author = {G. Almeida and others},
  title = {Universality, robustness, and limits of the eigenstate thermalization hypothesis in open quantum systems},
  journal = {Phys. Rev. E},
  volume = {113},
  year = {2026},
  number = {1},
  pages = {L012101},
  doi = {10.1103/qy19-nc4r}
}

@article{Shirai:2020,
  author = {T. Shirai and T. Mori},
  title = {Thermalization in open many-body systems based on eigenstate thermalization hypothesis},
  journal = {Phys. Rev. E},
  volume = {101},
  year = {2020},
  number = {4},
  pages = {042116},
  doi = {10.1103/PhysRevE.101.042116}
}

@article{Jahnke:2018off,
  author = {V. Jahnke},
  title = {Recent developments in the holographic description of quantum chaos},
  journal = {Adv. High Energy Phys.},
  volume = {2019},
  year = {2019},
  pages = {9632708},
  doi = {10.1155/2019/9632708},
  archiveprefix = {arXiv},
  eprint = {1811.06949},
  primaryclass = {hep-th}
}

@article{Evnin:2018jbh,
  author = {O. Evnin and W. Piensuk},
  title = {Quantum resonant systems, integrable and chaotic},
  journal = {J. Phys. A},
  volume = {52},
  year = {2019},
  number = {2},
  pages = {025102},
  doi = {10.1088/1751-8121/aaf2a1},
  archiveprefix = {arXiv},
  eprint = {1808.09173},
  primaryclass = {math-ph}
}

@article{Joel:2013,
  author = {K. Joel and D. Kollmar and L. Santos},
  title = {An introduction to the spectrum, symmetries, and dynamics of spin-1/2 Heisenberg chains},
  journal = {Am. J. Phys.},
  volume = {81},
  year = {2013},
  number = {6},
  pages = {450--457},
  doi = {10.1119/1.4798343}
}

@article{Dymarsky2018,
    author = "Dymarsky, Anatoly and Lashkari, Nima and Liu, Hong",
    title = "{Subsystem ETH}",
    eprint = "1611.08764",
    archivePrefix = "arXiv",
    primaryClass = "cond-mat.stat-mech",
    doi = "10.1103/PhysRevE.97.012140",
    journal = "Phys. Rev. E",
    volume = "97",
    pages = "012140",
    year = "2018"
}

@article{Beugeling:2015,
  author = {W. Beugeling and R. Moessner and M. Haque},
  title = {Off-diagonal matrix elements of local operators in many-body quantum systems},
  journal = {Phys. Rev. E},
  volume = {91},
  pages = {012144},
  year = {2015},
  doi = {10.1103/PhysRevE.91.012144},
  archiveprefix = {arXiv},
  eprint = {1407.2043},
  primaryclass = {cond-mat.stat-mech}
}

@article{Mondaini:2017,
  author = {R. Mondaini and M. Rigol},
  title = {Eigenstate thermalization in the two-dimensional transverse field {Ising} model. {II}. Off-diagonal matrix elements of observables},
  journal = {Phys. Rev. E},
  volume = {96},
  pages = {012157},
  year = {2017},
  doi = {10.1103/PhysRevE.96.012157},
  archiveprefix = {arXiv},
  eprint = {1705.08058},
  primaryclass = {cond-mat.stat-mech}
}

@article{Sala2020,
  author  = {Sala, Pablo and Rakovszky, Tibor and Verresen, Ruben and Knap, Michael and Pollmann, Frank},
  title   = {Ergodicity Breaking Arising from Hilbert Space Fragmentation in Dipole-Conserving Hamiltonians},
  journal = {Physical Review X},
  volume  = {10},
  pages   = {011047},
  year    = {2020},
  doi     = {10.1103/PhysRevX.10.011047}
}

@book{Haake:2010,
  author = {F. Haake},
  title = {Quantum Signatures of Chaos},
  edition = {3rd},
  publisher = {Springer},
  address = {Berlin},
  year = {2010},
  doi = {10.1007/978-3-642-05428-0}
}

@book{Stockmann:1999,
  author = {H.-J. St{\"o}ckmann},
  title = {Quantum Chaos: An Introduction},
  publisher = {Cambridge University Press},
  address = {Cambridge},
  year = {1999},
  doi = {10.1017/CBO9780511524622}
}

\end{document}